\documentclass[epj,nopacs]{svjour}
\usepackage{epsfig}
\usepackage{latexsym}
\usepackage{graphics}
\usepackage{amssymb}

\newcommand{\be}{\begin{equation}}
\newcommand{\ee}{\end{equation}}
\newcommand{\nn}{\nonumber}
\newcommand{\bea}{\begin{eqnarray}}
\newcommand{\eea}{\end{eqnarray}}

\journalname{EPJ C}
\begin{document}
\setcounter{page}{1}
\title{
  \vspace{-23pt}
  {\rightline{\small LU 2001/027, LPNHE 2001-14}}
 New method for calculating helicity amplitudes
 of jet--like QED processes for high--energy colliders \\
 I. Bremsstrahlung processes}
\titlerunning{New method for calculating helicity amplitudes}
\authorrunning{C.~Carimalo et al.}
\author{C.~Carimalo \inst{1}\thanks{carimalo@in2p3.fr} %
\and A.~Schiller \inst{2,}\thanks{Arwed.Schiller@itp.uni-leipzig.de} %
\and V.G.~Serbo \inst{3,}\thanks{serbo@math.nsc.ru}}
\institute{LPNHE, IN2P3-CNRS, Universit\'e Paris VI, F-75252 Paris, France 
\and 
Institut f\"ur Theoretische Physik and NTZ, Universit\"at Leipzig, D-04109
Leipzig, Germany 
\and 
Novosibirsk State University, Novosibirsk, 630090 Russia} 
\date{Received: December 19, 2001}

\abstract{
  Inelastic QED processes, the cross sections of which do not drop
  with increasing energy, play an important role at high--energy
  colliders. Such reactions have the form of two--jet processes
  with the exchange of a virtual photon in the $t$-channel. We
  consider them in the region of small scattering angles $ m/E \lesssim
  \theta \ll 1$, which yields the dominant contribution to their
  total cross sections.
  A new effective method is presented and applied to QED processes with
  emission of real photons to calculate the helicity amplitudes of these
  processes. Its basic idea is similar
  to the well--known equivalent--lepton method. Compact
  analytical expressions for those amplitudes up to $e^8$ are derived
  omitting only terms of the order of $m^2/E^2,\, \theta^2$, $\theta m/E$ 
  and higher order. The helicity amplitudes are presented in a
  compact  form in which
  large compensating terms are already cancelled.
  Some common properties for all jet--like processes are found and 
  we discuss
  their origin.}

\maketitle

\section{Introduction}

\subsection{Subject of the paper}

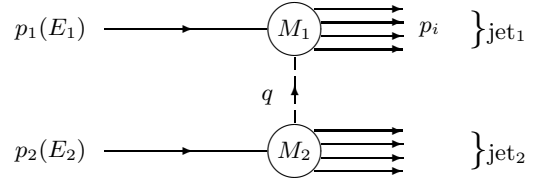
\begin{figure}[!htb]
  \centering
  \unitlength=1.8mm
  \begin{picture}(38.00,14.00)
    \put(23.00,12.00){\circle{4.00}}
    \put(23.00,12.00){\makebox(0,0)[cc]{$M_1$}}
    \put(23.00,3.00){\circle{4.00}}
    \put(23.00,3.00){\makebox(0,0)[cc]{$M_2$}}
    \put(9.00,3.00){\vector(1,0){6.50}}
    \put(9.00,12.00){\vector(1,0){6.50}}
    \put(14.00,3.00){\line(1,0){6.90}}
    \put(14.00,12.00){\line(1,0){6.90}}
    \put(25.00,12.50){\vector(1,0){6.00}}
    \put(25.00,11.50){\vector(1,0){6.00}}
    \put(24.50,13.50){\vector(1,0){6.50}}
    \put(23.00,5.10){\line(0,1){0.90}}
    \put(23.00,6.50){\line(0,1){2.00}}
    \put(23.00,6.50){\vector(0,1){1.40}}
    \put(23.00,9.00){\line(0,1){0.90}}
    \put(24.50,10.50){\vector(1,0){6.50}}
    \put(25.00,3.50){\vector(1,0){6.00}}
    \put(25.00,2.50){\vector(1,0){6.00}}
    \put(24.50,4.50){\vector(1,0){6.50}}
    \put(24.50,1.50){\vector(1,0){6.50}}
    \put(33.00,12.00){\makebox(0,0)[cc]{$p_i$}}
    \put(21.00,7.00){\makebox(0,0)[cc]{$q$}}
    \put(5.00,12.00){\makebox(0,0)[cc]{$p_1(E_1)$}}
    \put(5.00,3.00){\makebox(0,0)[cc]{$p_2(E_2)$}}
    \put(38.00,12.00){\makebox(0,0)[cc]{{\Large\}}jet$_1$}}
    \put(38.00,3.00){\makebox(0,0)[cc]{{\Large\}}jet$_2$}}
  \end{picture}
  \caption{ Generic block diagram of the two--jet process  $e e \to
  {\mathrm {jet}}_1 \ {\mathrm{jet}}_2$.}
  \label{fig:1}
\end{figure}

Accelerators with high-energy colliding $e^+e^-$, $\gamma e$,
$\gamma \gamma$ and $\mu^+ \mu^-$ beams are now widely used or
designed to study fundamental interactions~\cite{NLC}. Some
processes of quantum electrodynamics (QED) might play an
important role at these colliders, especially those inelastic
processes the cross section of which do not drop with increasing
energy. For this reason and since, in principle, the planned
colliders will be able to work with polarized particles, these
QED processes are required to be described in full detail,
including the calculation of their amplitudes with definite
helicities of all initial and final particles --- leptons ($l=e$
or $\mu$) and photons $\gamma$. These reactions have the form of
a two--jet process with the exchange of a virtual photon
$\gamma^*$ in the $t$-channel (Fig.~\ref{fig:1}).

We define by a jet kinematics in QED a high--energy reaction in
which the outgoing particles (leptons and photons) are produced
within a small cone $\theta_i \ll 1 $ relative to the
propagation axis of their respective parental incoming particle.
We work in the collider reference frame in which the initial
particles with 4-momenta $p_1$ and $p_2$ perform a head--on
collision with respective energies $E_1$ and $E_2$ of the same
order. The subject of our consideration is the jet--like process
of Fig.~\ref{fig:1} at high energies ($m_i$ is a lepton mass)
 \be
  s=2p_1 p_2 =4E_1 E_2 \gg m_i^2
  \label{1}
\ee
for arbitrary  helicities of leptons $\lambda_i=\pm 1/2$ and
photons $\Lambda_i=\pm 1$. The emission and scattering angles
$\theta_i$ are allowed to be much smaller than unity though they
may be of the order of the typical emission angles $m_i / E_i$
or larger. Stated differently, the transverse momenta of final
particles ${\bf p}_{i\perp}$ are allowed to be of the order of
the lepton mass or larger:
 \be
  \frac{m_i}{E_i} \lesssim \theta_i  \ll 1\,, \;\; m_i \lesssim
  |{\bf p}_{i\perp}| \ll E_i \,.
  \label{2}
\ee

The processes under discussion have large total cross sections.
Therefore, they present an essential background and they
determine particle losses in the beams and the beam life time.
Since all these processes can be calculated with high accuracy
independently of any model of strong interaction, they can
usefully serve for monitoring luminosity and polarization of
colliding beams. Besides, there is a specific feature exhibited
by some of jet--like processes --- the so-called MD or
beam--size effect (see review~\cite{KSS92} for detail).

All these properties of the jet--like QED processes justify the
growing interest to them from both the experimental and
theoretical communities in high--energy physics. Particular
problems related to these processes were discussed in a number
of original papers~\cite{Racah}--\cite{KSSSh00} and in
reviews such as \cite{BGMS,BFKK,KSS92}. But only recently (see
Ref.~\cite{KSSSh00}) the highly accurate analytical calculation
of the helicity amplitudes of all jet--like processes up to
$e^4$ (shown in Figs.~\ref{fig:2}--\ref{fig:10}) has been
completed.

In the above--mentioned original papers different approaches and
notations have been used. Here we develop a new simple and
effective method to calculate jet--like processes and apply it
to QED processes with emission of real photons. In particular,
we consider in detail the case of emission of up to three
photons along the direction of the initial particle
(Fig.~\ref{fig:11}). 

It is quite important to realize that at high energies (\ref{1})
the region of scattering angles (\ref{2}) gives the dominant
contribution to the cross sections of all QED jet--like
processes such as those shown in
Figs.~\ref{fig:2}--\ref{fig:11}. Just in this region the
amplitudes of these processes can be found in a ``final form''
including the polarizations of all particles. By this we mean
that we obtain compact and simple analytical expressions for all
helicity amplitudes with high accuracy, omitting only terms of
the order of
 \be
  \frac{m_i^2}{E_i^2} \ , \ \ \theta_i^2 \ , \ \
  \theta_i\cdot\frac{m_i}{E_i}
  \label{3}
 \ee
or higher order. Namely, the amplitude $M_{fi}$ of any process given in
Figs.~\ref{fig:1}--\ref{fig:11} can be presented in a simple
factorized form
\be
  M_{fi}= \frac{s}{q^2} J_1 J_2
  \label{4}
\ee
where the impact factors $J_1$  and $J_2$ do not depend on $s$. The
impact factor $J_1$  corresponds to the first jet (or the upper
block) and the impact factor $J_2$ corresponds to the second jet (or
the lower block) of Fig.~\ref{fig:1}.

We give analytical expressions for these impact factors. They
are not only compact but are also very convenient for numerical
calculations. The reason is that we present the impact factors
in such a form that large compensating terms are already
cancelled. It is well known that this problem of large
compensating terms is very difficult to manage in all computer
packages like CompHEP \cite{CompHEP} which generate
automatically matrix elements and compute cross sections.

It should be noted that the discussed approximation differs
considerably from the known approach of the CALCUL group and
others \cite{CALCUL} where such processes are calculated for not
too small scattering angles $\theta_i \gg m_i/E_i$. In that
approach terms of the order of $m_i/|{\bf p}_{i\perp}|$ are
neglected, which, however, may give the dominant contribution to
the total cross sections.

To get the high--energy helicity amplitudes in the jet--like
kinematics as compact analytical expressions and to make the
calculations very efficient we systematically exploit three
basic ideas: (i) a convenient decomposition of all 4--momenta
into large and small components (using the
so--called Sudakov or light--cone variables); (ii) gauge
invariance of the amplitudes is used in order to combine large
terms into finite expressions; (iii) the calculations are
considerably simplified in replacing the numerators of lepton
propagators by vertices involving real leptons and antileptons.
All these ideas are not new. In particular, the last one is the
basis of the equivalent--lepton
method~\cite{Kessler60,BFKhquasi,GSquasi} and has been used to
calculate some QCD amplitudes with massless quarks~\cite{Farra}.
However, as we will demonstrate in the present paper, the
combination of these ideas leads to a very efficient way in
calculating the amplitudes of interest in the jet kinematics 
here considered.

\subsection{Jet--like QED processes up
to fourth order with nondecreasing cross sections}

We consider electromagnetic interactions  of electrons,
positrons and photons on tree-level in high--energy $ee$,
$e\gamma$ and $\gamma\gamma$ collisions.
To third and fourth orders in the electromagnetic coupling $e$
the corresponding jet--like QED processes in the form of block
diagrams are shown in Figs.~\ref{fig:2}--\ref{fig:10}. Solid
lines represent leptons, dashed one photons. Only those diagrams
are drawn that give the dominant contributions at high energies.
\begin{figure}[!htb]
  \begin{center}
  \setlength{\unitlength}{1.0cm}
    \begin{picture}(6.0,2.5)
      \unitlength=2.00mm
      \put(12.00,12.00){\circle{3.40}}
      \put( 0.00, 0.00){\line(1,0){12.00}}
      \put( 0.00, 0.00){\vector(1,0){6.4}}
      \put( 0.00,10.90){\line(1,0){10.60}}
      \put( 0.00,10.90){\vector(1,0){6.4}}
      \put(12.00, 0.00){\line(0,1){2.0}}
      \put(12.00, 2.50){\line(0,1){2.0}}
      \put(12.00, 5.00){\line(0,1){2.0}}
      \put(12.00, 5.00){\vector(0,1){1.40}}
      \put(12.00, 7.50){\line(0,1){2.0}}
      \put(12.00,10.00){\line(0,1){0.2}}
      \put(12.00, 0.00){\vector(1,0){6.4}}
      \put(12.00, 0.00){\line(1,0){12.0}}
      \put(13.40,13.10){\line(1,0){0.6}}
      \put(14.50,13.10){\line(1,0){2.0}}
      \put(17.00,13.10){\line(1,0){2.0}}
      \put(17.00,13.10){\vector(1,0){1.40}}
      \put(19.50,13.10){\line(1,0){2.0}}
      \put(22.00,13.10){\line(1,0){2.0}}
      \put(13.40,10.90){\vector(1,0){5.0}}
      \put(13.40,10.90){\line(1,0){10.6}}
    \end{picture}
  \end{center}
  \caption{Single bremsstrahlung in $e e$ collisions:
             $e e \to e e \gamma$.}
  \label{fig:2}
\end{figure}
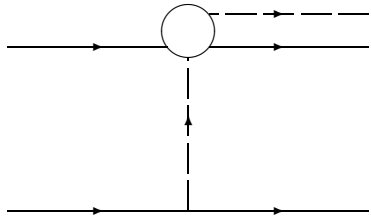
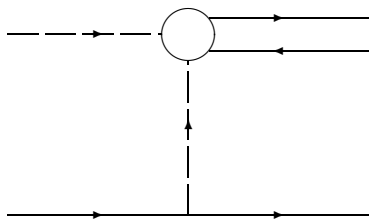
\begin{figure}[!htb]
  \begin{center}
  \setlength{\unitlength}{1cm}
    \begin{picture}(6.0,2.0)
      \unitlength=2.00mm
      \put(12.00,12.00){\circle{3.40}}
      \put( 0.00, 0.00){\line(1,0){12.00}}
      \put( 0.00, 0.00){\vector(1,0){6.4}}
      \put( 0.00,12.00){\line(1,0){2.0}}
      \put( 2.50,12.00){\line(1,0){2.0}}
      \put( 5.00,12.00){\line(1,0){2.0}}
      \put( 5.00,12.00){\vector(1,0){1.40}}
      \put( 7.50,12.00){\line(1,0){2.0}}
      \put(10.00,12.00){\line(1,0){0.2}}
      \put(12.00, 0.00){\line(0,1){2.0}}
      \put(12.00, 2.50){\line(0,1){2.0}}
      \put(12.00, 5.00){\line(0,1){2.0}}
      \put(12.00, 5.00){\vector(0,1){1.40}}
      \put(12.00, 7.50){\line(0,1){2.0}}
      \put(12.00,10.00){\line(0,1){0.2}}
      \put(12.00, 0.00){\vector(1,0){6.4}}
      \put(12.00, 0.00){\line(1,0){12.0}}
      \put(13.40,10.90){\line(1,0){10.6}}
      \put(13.40,13.10){\vector(1,0) {5.00}}
      \put(24.00,10.90){\vector(-1,0){6.40}}
      \put(13.40,13.10){\line(1,0){10.60}}
    \end{picture}
  \end{center}
    \caption{Single lepton pair production in $\gamma e$
             collisions: $\gamma e \to l^+ l^- e$.}
    \label{fig:3}
\end{figure}
\begin{figure}[!htb]
  \begin{center}
  \setlength{\unitlength}{1cm}
  \begin{picture}(6.0,3.0)
      \unitlength=2.00mm
      \put(12.00, 0.00){\circle{3.40}}
      \put(12.00,12.00){\circle{3.40}}
      \put( 0.00, 1.10){\line(1,0){10.60}}
      \put( 0.00, 1.10){\vector(1,0){6.4}}
      \put( 0.00,10.90){\line(1,0){10.60}}
      \put( 0.00,10.90){\vector(1,0){6.4}}
      \put(12.00, 1.70){\line(0,1){0.2}}
      \put(12.00, 2.50){\line(0,1){2.0}}
      \put(12.00, 5.00){\line(0,1){2.0}}
      \put(12.00, 5.00){\vector(0,1){1.40}}
      \put(12.00, 7.50){\line(0,1){2.0}}
      \put(12.00,10.00){\line(0,1){0.2}}
      \put(13.40,-1.10){\line(1,0){0.6}}
      \put(14.50,-1.10){\line(1,0){2.0}}
      \put(17.00,-1.10){\line(1,0){2.0}}
      \put(17.00,-1.10){\vector(1,0){1.40}}
      \put(19.50,-1.10){\line(1,0){2.0}}
      \put(22.00,-1.10){\line(1,0){2.0}}
      \put(13.40, 1.10){\vector(1,0){5.0}}
      \put(13.40, 1.10){\line(1,0){10.6}}
      \put(13.40,13.10){\line(1,0){0.6}}
      \put(14.50,13.10){\line(1,0){2.0}}
      \put(17.00,13.10){\line(1,0){2.0}}
      \put(17.00,13.10){\vector(1,0){1.40}}
      \put(19.50,13.10){\line(1,0){2.0}}
      \put(22.00,13.10){\line(1,0){2.0}}
      \put(13.40,10.90){\vector(1,0){5.0}}
      \put(13.40,10.90){\line(1,0){10.6}}
     \end{picture}
  \end{center}
     \caption{Double bremsstrahlung with single photons
     along each initial
              lepton direction: $e e \to e e \gamma \gamma$.}
     \label{fig:4}
  \end{figure}
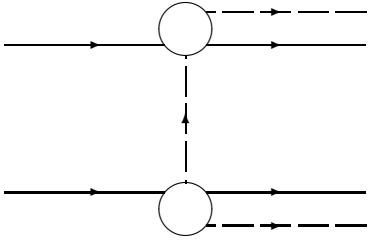
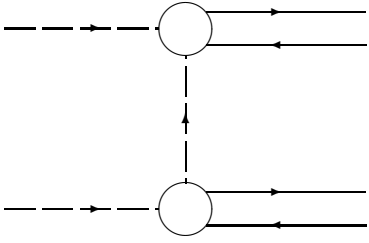
\begin{figure}[!htb]
  \begin{center}
    \setlength{\unitlength}{1cm}
    \begin{picture}(6.0,2.5)
      \unitlength=2.00mm
      \put(12.00, 0.00){\circle{3.40}}
      \put(12.00,12.00){\circle{3.40}}
      \put( 0.00, 0.00){\line(1,0){2.0}}
      \put( 2.50, 0.00){\line(1,0){2.0}}
      \put( 5.00, 0.00){\line(1,0){2.0}}
      \put( 5.00, 0.00){\vector(1,0){1.40}}
      \put( 7.50, 0.00){\line(1,0){2.0}}
      \put(10.00, 0.00){\line(1,0){0.2}}
      \put( 0.00,12.00){\line(1,0){2.0}}
      \put( 2.50,12.00){\line(1,0){2.0}}
      \put( 5.00,12.00){\line(1,0){2.0}}
      \put( 5.00,12.00){\vector(1,0){1.40}}
      \put( 7.50,12.00){\line(1,0){2.0}}
      \put(10.00,12.00){\line(1,0){0.2}}
      \put(12.00, 1.70){\line(0,1){0.3}}
      \put(12.00, 2.50){\line(0,1){2.0}}
      \put(12.00, 5.00){\line(0,1){2.0}}
      \put(12.00, 5.00){\vector(0,1){1.40}}
      \put(12.00, 7.50){\line(0,1){2.0}}
      \put(12.00,10.00){\line(0,1){0.2}}
      \put(13.40,-1.10){\line(1,0){10.6}}
      \put(24.00,-1.10){\vector(-1,0){6.40}}
      \put(13.40, 1.10){\vector( 1,0){5.00}}
      \put(13.40, 1.10){\line(1,0){10.60}}
      \put(13.40,10.90){\line(1,0){10.6}}
      \put(24.00,10.90){\vector(-1,0){6.40}}
      \put(13.40,13.10){\vector( 1,0){5.00}}
      \put(13.40,13.10){\line(1,0){10.60}}
    \end{picture}
    \end{center}
    \caption{Double lepton pair production in $\gamma \gamma$
             collisions: $\gamma \gamma \to e^+ e^- l^+ l^- $.}
    \label{fig:5}
\end{figure}
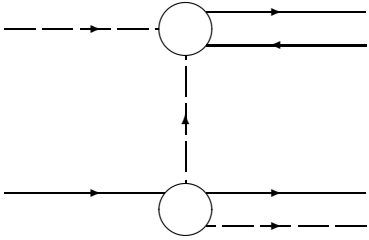
\begin{figure}[!htb]
  \begin{center}
  \setlength{\unitlength}{1cm}
    \begin{picture}(6.0,2.5)
       \unitlength=2.00mm
       \put(12.00, 0.00){\circle{3.40}}
       \put(12.00,12.00){\circle{3.40}}
       \put( 0.00, 1.10){\line(1,0){10.60}}
       \put( 0.00, 1.10){\vector(1,0){6.4}}
       \put( 0.00,12.00){\line(1,0){2.0}}
       \put( 2.50,12.00){\line(1,0){2.0}}
       \put( 5.00,12.00){\line(1,0){2.0}}
       \put( 5.00,12.00){\vector(1,0){1.40}}
       \put( 7.50,12.00){\line(1,0){2.0}}
       \put(10.00,12.00){\line(1,0){0.2}}
       \put(12.00, 1.70){\line(0,1){0.2}}
       \put(12.00, 2.50){\line(0,1){2.0}}
       \put(12.00, 5.00){\line(0,1){2.0}}
       \put(12.00, 5.00){\vector(0,1){1.40}}
       \put(12.00, 7.50){\line(0,1){2.0}}
       \put(12.00,10.00){\line(0,1){0.2}}
       \put(13.40,-1.10){\line(1,0){0.6}}
       \put(14.50,-1.10){\line(1,0){2.0}}
       \put(17.00,-1.10){\line(1,0){2.0}}
       \put(17.00,-1.10){\vector(1,0){1.40}}
       \put(19.50,-1.10){\line(1,0){2.0}}
       \put(22.00,-1.10){\line(1,0){2.0}}
       \put(13.40, 1.10){\vector(1,0){5.0}}
       \put(13.40, 1.10){\line(1,0){10.6}}
       \put(13.40,10.90){\line(1,0){10.6}}
       \put(13.40,13.10){\vector(1,0) {5.00}}
       \put(24.00,10.90){\vector(-1,0){6.40}}
       \put(13.40,13.10){\line(1,0){10.60}}
    \end{picture}
    \end{center} 
    \caption{Process  $\gamma e \to l^+ l^- e \gamma$ with a
    final photon along the initial lepton direction.}
    \label{fig:6}
\end{figure}
\begin{figure}[!htb]
  \begin{center}
  \setlength{\unitlength}{1cm}
    \begin{picture}(6.0,2.5)
       \unitlength=2.00mm
       \put(12.00, 6.00){\circle{3.40}}
       \put(13.40, 4.90){\line(1,0){10.6}}
       \put(13.40, 7.10){\vector(1,0) {5.00}}
       \put(24.00, 4.90){\vector(-1,0){6.40}}
       \put(13.40, 7.10){\line(1,0){10.60}}
       \put( 0.00, 0.00){\line(1,0){12.00}}
       \put( 0.00, 0.00){\vector(1,0){6.4}}
       \put( 0.00,12.00){\line(1,0){12.00}}
       \put( 0.00,12.00){\vector(1,0){6.4}}
       \put(12.00, 0.00){\line(0,1){2.0}}
       \put(12.00, 0.00){\vector(0,1){1.40}}
       \put(12.00, 2.50){\line(0,1){1.7}}
       \put(12.00, 7.80){\line(0,1){1.7}}
       \put(12.00,10.00){\line(0,1){2.0}}
       \put(12.00,12.00){\vector(0,-1){1.40}}
       \put(12.00, 0.00){\vector(1,0){6.4}}
       \put(12.00, 0.00){\line(1,0){12.0}}
       \put(12.00,12.00){\vector(1,0){6.4}}
       \put(12.00,12.00){\line(1,0){12.0}}
       %
    \end{picture}
  \end{center}
    \caption{Two--photon pair production in $e e$ collisions:
             $e e \to e e \; l^-l^+ $.}
    \label{fig:7}
\end{figure}
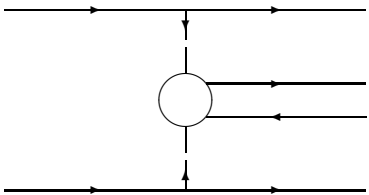
\begin{figure}[!htb]
  \begin{center}
  \setlength{\unitlength}{1cm}
    \begin{picture}(6.0,2.5)
       \unitlength=2.00mm
       \put(12.00,12.00){\circle{3.40}}
       \put( 0.00, 0.00){\line(1,0){12.00}}
       \put( 0.00, 0.00){\vector(1,0){6.4}}
       \put( 0.00,10.90){\line(1,0){10.60}}
       \put( 0.00,10.90){\vector(1,0){6.4}}
       \put(12.00, 0.00){\line(0,1){2.0}}
       \put(12.00, 2.50){\line(0,1){2.0}}
       \put(12.00, 5.00){\line(0,1){2.0}}
       \put(12.00, 5.00){\vector(0,1){1.40}}
       \put(12.00, 7.50){\line(0,1){2.0}}
       \put(12.00,10.00){\line(0,1){0.2}}
       \put(12.00, 0.00){\vector(1,0){6.4}}
       \put(12.00, 0.00){\line(1,0){12.0}}
       \put(13.40,13.10){\line(1,0){0.6}}
       \put(14.50,13.10){\line(1,0){2.0}}
       \put(17.00,13.10){\line(1,0){2.0}}
       \put(14.50,13.10){\vector(1,0){1.40}}
       \put(19.00,13.10){\line(4,1){5.0}}
       \put(19.00,13.10){\vector(4,1){2.9}}
       \put(19.00,13.10){\line(4,-1){5.0}}
       \put(24.00,11.85){\vector(-4, 1){2.9}}
       \put(13.40,10.90){\vector(1,0){5.0}}
       \put(13.40,10.90){\line(1,0){10.6}}
       %
    \end{picture}
  \end{center}
    \caption{Bremsstrahlung pair production in $ee$ collisions:
             $e e \to e e \; l^+ l^-$.}
    \label{fig:8}
\end{figure}
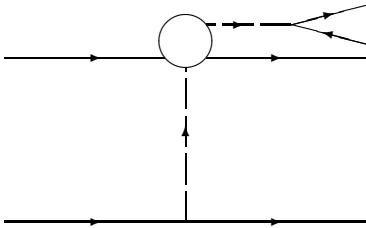
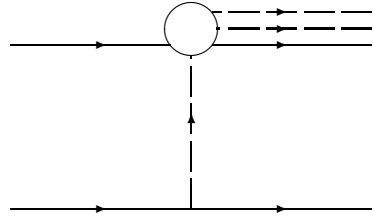
\begin{figure}[!htb]
  \begin{center}
  \setlength{\unitlength}{1cm}
    \begin{picture}(6.0,3.0)
      \unitlength=2.00mm
      \put(12.00,12.00){\circle{3.40}}
      \put( 0.00, 0.00){\line(1,0){12.00}}
      \put( 0.00, 0.00){\vector(1,0){6.4}}
      \put( 0.00,10.90){\line(1,0){10.60}}
      \put( 0.00,10.90){\vector(1,0){6.4}}
      \put(12.00, 0.00){\line(0,1){2.0}}
      \put(12.00, 2.50){\line(0,1){2.0}}
      \put(12.00, 5.00){\line(0,1){2.0}}
      \put(12.00, 5.00){\vector(0,1){1.40}}
      \put(12.00, 7.50){\line(0,1){2.0}}
      \put(12.00,10.00){\line(0,1){0.2}}
      \put(12.00, 0.00){\vector(1,0){6.4}}
      \put(12.00, 0.00){\line(1,0){12.0}}
      \put(13.40,13.10){\line(1,0){0.6}}
      \put(14.50,13.10){\line(1,0){2.0}}
      \put(17.00,13.10){\line(1,0){2.0}}
      \put(17.00,13.10){\vector(1,0){1.40}}
      \put(19.50,13.10){\line(1,0){2.0}}
      \put(22.00,13.10){\line(1,0){2.0}}
      \put(13.70,12.00){\line(1,0){0.2}}
      \put(14.50,12.00){\line(1,0){2.0}}
      \put(17.00,12.00){\line(1,0){2.0}}
      \put(17.00,12.00){\vector(1,0){1.40}}
      \put(19.50,12.00){\line(1,0){2.0}}
      \put(22.00,12.00){\line(1,0){2.0}}
      \put(13.40,10.90){\vector(1,0){5.0}}
      \put(13.40,10.90){\line(1,0){10.6}}
    \end{picture}
  \end{center}
    \caption{Double bremsstrahlung  $e e \to e e \gamma \gamma$
               with two photons along the direction of one
                initial lepton.}
    \label{fig:9}
\end{figure}
\begin{figure}[!htb]
  \begin{center}
  \setlength{\unitlength}{1cm}
    \begin{picture}(6.0,3.0)
      \unitlength=2.00mm
      \put(12.00,12.00){\circle{3.40}}
      \put( 0.00, 0.00){\line(1,0){12.00}}
      \put( 0.00, 0.00){\vector(1,0){6.4}}
      \put( 0.00,10.90){\line(1,0){2.0}}
      \put( 2.50,10.90){\line(1,0){2.0}}
      \put( 6.50,10.90){\line(1,0){2.0}}
      \put( 9.00,10.90){\line(1,0){1.60}}
      \put( 5.00,10.90){\vector(1,0){1.0}}
      \put(12.00, 0.00){\line(0,1){2.0}}
      \put(12.00, 2.50){\line(0,1){2.0}}
      \put(12.00, 5.00){\line(0,1){2.0}}
      \put(12.00, 5.00){\vector(0,1){1.40}}
      \put(12.00, 7.50){\line(0,1){2.0}}
      \put(12.00,10.00){\line(0,1){0.2}}
      \put(12.00, 0.00){\vector(1,0){6.4}}
      \put(12.00, 0.00){\line(1,0){12.0}}
      \put(13.40,10.90){\line(1,0){0.6}}
      \put(14.50,10.90){\line(1,0){2.0}}
      \put(17.00,10.90){\line(1,0){2.0}}
      \put(17.00,10.90){\vector(1,0){1.40}}
      \put(19.50,10.90){\line(1,0){2.0}}
      \put(22.00,10.90){\line(1,0){2.0}}
      \put(13.70,12.00){\line(1,0){10.3}}
      \put(19.80,12.00){\vector(-1,0){1.40}}
      %
      \put(13.40,13.10){\vector(1,0){5.0}}
      \put(13.40,13.10){\line(1,0){10.6}}
    \end{picture}
  \end{center}
    \caption{Process  $\gamma e \to \gamma l^+l^- e$ with
    the both final photon and lepton pair along the direction
    of the initial photon.}
    \label{fig:10}
\end{figure}
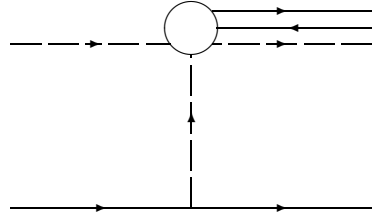
\begin{figure}[!htb]
  \begin{center}
  \setlength{\unitlength}{1cm}
    \begin{picture}(6.0,3.0)
      \unitlength=2.00mm
      \put(12.00,12.55){\circle{4.80}}
      \put( 0.00, 0.00){\line(1,0){12.00}}
      \put( 0.00, 0.00){\vector(1,0){6.4}}
      \put( 0.00,10.90){\line(1,0){10.00}}
       \put(0.00,10.90){\vector(1,0){6.4}}
      \put(12.00, 0.00){\line(0,1){2.0}}
      \put(12.00, 2.50){\line(0,1){2.0}}
      \put(12.00, 5.00){\line(0,1){2.0}}
      \put(12.00, 5.00){\vector(0,1){1.40}}
      \put(12.00, 7.50){\line(0,1){2.0}}
      \put(12.00,10.00){\line(0,1){0.2}}
      \put(12.00, 0.00){\vector(1,0){6.4}}
      \put(12.00, 0.00){\line(1,0){12.0}}
      %
      \put(14.70,13.10){\line(1,0){1.8}}
      \put(17.00,13.10){\line(1,0){2.0}}
      \put(17.00,13.10){\vector(1,0){1.40}}
      \put(19.50,13.10){\line(1,0){2.0}}
      \put(22.00,13.10){\line(1,0){2.0}}
      \put(14.00,14.20){\line(1,0){0.2}}
      \put(14.70,14.20){\line(1,0){1.8}}
      \put(17.00,14.20){\line(1,0){2.0}}
      \put(17.00,14.20){\vector(1,0){1.40}}
      \put(19.50,14.20){\line(1,0){2.0}}
      \put(22.00,14.20){\line(1,0){2.0}}

      \put(14.70,12.00){\line(1,0){1.8}}
      \put(17.00,12.00){\line(1,0){2.0}}
      \put(17.00,12.00){\vector(1,0){1.40}}
      \put(19.50,12.00){\line(1,0){2.0}}
      \put(22.00,12.00){\line(1,0){2.0}}
      \put(14.00,10.90){\vector(1,0){4.4}}
      \put(14.00,10.90){\line(1,0){10.0}}
    \end{picture}
  \end{center}
    \caption{Triple bremsstrahlung  $e e \to e e
    \gamma \gamma \gamma$
    with three photons along the direction of one
    initial lepton.}
    \label{fig:11}
\end{figure}
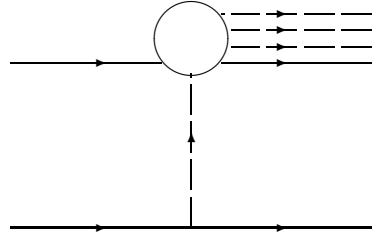

The third order processes are: single bremsstrahlung in $e^{\pm}
e$ collision (Fig.~\ref{fig:2}) and single lepton pair $l^+ l^-$
production in $\gamma e$ collision (Fig.~\ref{fig:3}). To the
fourth order processes belong lepton pair production and
bremsstrahlung in various combinations, including simple
combinations of the above--mentioned third order processes
(Figs.~\ref{fig:4}--\ref{fig:6}) and new types of reactions
(Figs.~\ref{fig:7}--\ref{fig:10}).

The discussed processes are important for the following reasons:

1) 
Some of these reactions are used (or are proposed to be used)
as monitoring processes to determine the collider luminosity and
to measure the polarization of the colliding particles. For
example, the double bremsstrahlung process (Fig.~\ref{fig:4})
has been used as the standard calibration process at several
colliders in Novosibirsk, Frascati and Orsay
\cite{golubnichy,augustin}. In Ref.~\cite{leplumsag} it has been
suggested to use the single bremsstrahlung process
(Fig.~\ref{fig:2}) for measuring the luminosity and the
polarization of the initial $e^\pm$ at the LEP collider (see
also paper~\cite{kotkinpl89}). As it was demonstrated 
experimentally (see Ref.~\cite{leplum}), that process
has a  good chance to be used for luminosity purposes. The same
process has been proposed~\cite{courau} to measure the
luminosity at the DA$\Phi$NE collider (see also
Ref.~\cite{Carimalo}). The processes $\gamma \gamma \to \mu^+
\mu^- e^+ e^-$ and $\gamma \gamma \to \mu^+ \mu^- \mu^+ \mu^-$
of Fig.~\ref{fig:5} may be useful to monitor colliding $\gamma
\gamma$ beams \cite{beams,kuraevnp85,CKS}. Finally, the
possibility of designing $\mu^+ \mu^-$ colliders is widely
discussed at present. Therefore, the processes $\mu^+ \mu^- \to
l^+ l^- l^+l^- $ ($l=e,\mu$) may be useful for luminosity
measurements at those colliders~\cite{mumuGinz}. Recently, 
processes of Figs. \ref{fig:7} and \ref{fig:8} have been taken
into account as radiative corrections to the unpolarized Bhabha
scattering used as calibration process at LEP~\cite{recent}.

2)
Due to their large cross sections those reactions
contribute as significant background to a number of
experiments in the electroweak sector and to hadronic cross
sections. For example, the background process $e^+e^- \to e^+
e^- \mu^+ \mu^-$ is of special importance for experiments
studying two--photon and bremsstrahlung  production of 
$\pi^+\pi^-$ systems due to the known experimental difficulties
in discriminating pions and muons \cite{GSS2000}.

3) 
The bremsstrahlung process of Fig. \ref{fig:2} is of special
importance for storage rings since it is the leading beam loss
mechanism: after emitting a photon with energy above
approximately $1$ \%, the electron leaves the beam. Therefore,
the luminosity and the beam life time of the $e^+e^-$ storage
rings depends strongly on the properties of this 
reaction~\cite{Burkhard}.

4)
The methods to calculate the helicity amplitudes of
these processes and to obtain some distributions for the latter
can be easily translated to several semihard QCD processes such as
$\gamma \gamma \to q \bar{q} Q \bar{Q}$ \cite{kuraevnp85} ($q$
and $Q$ are different quarks) and $\gamma \gamma \to M M'$,
$\gamma \gamma \to M q \bar{q}$ \cite{mesons} ($M, M'$ denote
neutral mesons such as $\rho^0,\, \omega,\, \phi,\, \Psi,\,
\pi^0,\, a_2 ...$).

The outline of the paper is as follows. In Sect. 2 we describe
the method for the  calculation of the helicity amplitudes. In
the next section we derive all vertices necessary for the
bremsstrahlung processes and discuss their properties. Sections
4--6 are devoted to the calculation of the impact factors for
single, double and multiple bremsstrahlung. Some general
properties of the impact factors are discussed in Sect.~7. The
final chapter summarizes our results. 
In the Appendix we collect some
formulae for the Dirac bispinors and matrices in the so--called
spinor or chiral representation that appears to be useful
for calculation in the region of small angles.

\section{Method for calculation of helicity amplitudes }

\subsection{Sudakov or light-cone variables}

Let us introduce some notations using the block diagram of
Fig.~\ref{fig:1} for example. We use a collider reference frame
with the $z$--axis along the momentum ${\bf p}_1$, the azimuthal
angles are denoted by $\varphi_i$ (they are referred to one fixed
$x$-axis). It is convenient to introduce the light-like 4--vectors
$P_1$ and $P_2$:
\bea
  P_1&=&p_1 -{m_1^2\over s a}\, p_2,\;\; P_2=
  p_2 -{m_2^2\over s a}\, p_1,
  \;\;
  P_1^2 = {P}_2^2 = 0\,,
  \nn\\
  s&=&2 p_1 p_2, \;\; 
  \nn\\ 
  \tilde s&=&(P_1+P_2)^2 =
  2 P_1 P_2 = s \left( 1 -
  {4\epsilon\over a} + {\epsilon\over a^2} \right)\,,
  \label{5}\\
  a &=& {1\over 2}\left(1+ \sqrt{1-4\epsilon}\,\right)\approx
  1-\epsilon\,,\;\; \epsilon= {m_1^2 m_2^2\over s^2}
  \nn
\eea
and to decompose any 4--vector $A$ into components in the plane
spanned by the 4--vectors $P_1$ and $P_2$ and components in the
plane orthogonal to them
\bea
  &&A= x_A P_1 + y_A P_2+ A_\perp\,,\;\;
  A^2 = \tilde s x_A y_A + A_\perp^2 \,,
  \nn\\
  &&x_A=\frac{ 2 A P_2}
  {\tilde s}\,,\;\;y_A=\frac{ 2 A P_1} {\tilde s} \,.
  \label{6}
\eea
The parameters $x_A$ and $y_A$ are the so-called {\it
Sudakov variables} (they often are referred also as {\it
light-cone variables}). In the used reference frame
\begin{eqnarray*}
  &&  P_1=E_1 a_1\,(1,\,0,\,0,\,1)\,,\;\;
  P_2=E_2 a_2\,(1,\,0,\,0,\,-1)\,,
  \nn\\
  &&  a_i = 1-\frac{m_i^2}{E_i^2 a}\approx 1- \frac{m_i^2}{E_i^2}
\end{eqnarray*}
and the 4--vector $A_{\perp}$ has $x$ and $y$ components
only, e.g.
 $$
  A_\perp = (0,\, A_x ,\, A_y, \,0) = (0,\, {\bf A}_\perp ,\, 0) \,,
  \;\; \; A^2_\perp =- {\bf A}^2_\perp\, .
$$

Omitting terms of the order of $\epsilon$ only, we have
$$
  \tilde s =s\,,\;\; A^2= s x_A y_A - {\bf A}_\perp^2\,,\;\;
  x_A=\frac{ 2 A P_2} {s}\,,\;\;y_A=\frac{ 2 A P_1} {s}\,.
$$
For the colliding particles the Sudakov variables are
\be
  x_1=1\,,\;\; y_1= \frac{m_1^2}{s}\,,\;\; x_2=\frac{m_2^2}{s}\,,\;\;
  y_2=1\,.
  \label{7}
\ee
We also note the useful relation for all external momenta
\be
  p_i^2 = m_i^2 = s x_i y_i -{\bf p}^2 _{i\perp}
  \label{8}
 \ee
which means that for each external momentum only three
parameters are independent (say, $x_i$ and ${\bf p}_{i\perp}$
for the first jet).

The 4--vectors $p_i$ of particles from the first jet have large
components along $P_1$ and small ones along $P_2$. Therefore, in
the limit $s\to \infty$ [with accuracy (\ref{3})] the parameters
 \be
  x_i =\frac{2p_i P_2}{s} = \frac{E_i}{E_1}\,, \;\; i\in \;
  \mbox{jet}_1
  \label{9}
\ee
are finite, whereas
\be
  y_i =\frac{2p_i P_1}{s} = \frac{m^2_i +{\bf p}_{i\perp}^2}
  {s x_i}\,,\;\; i\in \;\;\mbox{jet}_1
  \label{10}
 \ee
are small. The Sudakov variable $x_i$ is  the fraction of energy
of the first incoming particle carried by the $i$-th final
particle. Analogously, for a 4--vector $p_l$ of  particles from
the second jet the parameters
 \be
  y_l =\frac{2p_l P_1}{s} = \frac{E_l}{E_2}\,, \;\; l\in \;\mbox{jet}_2
 \label{11}
\ee
are finite, whereas
\be
  x_l ={2p_l P_2 \over s} = {m^2_l +{\bf p}_{l\perp}^2 \over s y_l}\,,
  \;\; l\in \;\;\mbox{jet}_2
  \label{12}
 \ee
are small. The parameter $y_l$ is the fraction of energy of the
second initial particle carried by the $l$-th final particle. Since
 $$
  x_q = x_2 - \sum_{l\in {\rm jet}_2}\, x_l\,,\;\;\;
  y_q = \sum_{i\in {\rm jet}_1}\, y_i- y_1\,,
 $$
the Sudakov parameters for the virtual photon are small and,
therefore,
 \be
  \sum_{i\in {\rm jet}_1} x_i =x_1= 1\,,\;\; \sum_{l\in {\rm
  jet}_2} y_l=y_2= 1\,.
  \label{13}
\ee

Now we discuss the Sudakov decomposition of the photon
polarization vector, using a final photon from the first jet 
for example. We would like to remind that we have chosen  a
coordinate system with a fixed $x$-axis transverse to the beam
direction.
Let $e \equiv e^{(\Lambda)}(k)$ be the polarization 4-vector of
that photon with 4--momentum $k$ and helicity $\Lambda = \pm 1$.
Using gauge invariance, this vector can be replaced by $e+\zeta
k$. The arbitrary parameter $\zeta$ is chosen in such a way that
the new polarization vector (for which we use the same notation
$e$) has no a component along $P_1$, i.e.
 \be
  e=y_e P_2+{e}_\perp\,.
  \label{14}
 \ee
The parameter $y_e$ is determined from the condition $e k=0$:
 \be
  sy_e =\frac{-2 k_\perp {e}_\perp}{x_k}\,,\;\;
  x_k = \frac{2kP_2}{ s}\,.
  \label{15}
 \ee
Since $P_2P_2=P_2 e_\perp =0$, the transverse component
${e}_\perp$ satisfies the usual normalization condition
$$
  {e}_\perp^{(\Lambda)\,*}\,{e}_\perp^{(\Lambda')} =
  e^{(\Lambda)\,*}(k)\,e^{(\Lambda')}(k)=- \delta_{\Lambda
  \Lambda'}
$$
and can be chosen as
\be
  {e}_\perp\equiv {e}_\perp^{(\Lambda)} = -{\Lambda\over
  \sqrt{2}}\, (0,\,1,\,{\rm i} \Lambda ,\,0) =
  -{e}_\perp^{(-\Lambda)\,*}\,.
  \label{16}
 \ee
Therefore, ${e}_\perp$ does not depend on the 4--momentum of the
photon $k$ contrary to the polarization vector $e$ itself which
depends on $k$ via the parameter $y_e$. This, indeed, is very
convenient in the further calculations since we can choose the
same form of the transverse 4--vector $e_\perp$ for all final
photons in the first jet.
{}For the photon with helicity $\tilde\Lambda$ and polarization
vector $\tilde e =\tilde{x}_e P_1+\tilde{e}_\perp$ in the second
jet we use the relation
\be
  \tilde{e}_\perp^{(\tilde \Lambda)} = {e}_\perp^{(-\tilde\Lambda)}\,.
  \label{16a}
\ee

In the following we systematically neglect terms of the
relative order of (\ref{3}).

\subsection{Helicity amplitudes in factorized form}

The amplitude $M_{fi}$ corresponding to the diagram of
Fig.~\ref{fig:1} can be written in the form
 \be
  M_{fi} = M_1^{\mu} \; {g_{\mu \nu} \over q^2} \; M_2^\nu ,
  \label{17}
 \ee
where $M_1^\mu$ and $M_2^\nu$ are the amplitudes of the upper
and lower block of Fig.~\ref{fig:1}, respectively, and
$g_{\mu\nu}$ is the metric tensor. The transition amplitude
$M_1$ describes the scattering of an incoming particle of
momentum $p_1$ with a virtual photon and subsequent transition
to the first jet (similar for $M_2$).

We will show now that the amplitude of the process can be
presented  with accuracy (\ref{3}) in factorized form:
\bea
  M_{fi} &=& {s\over q^2}\, J_1 \, J_2\,,
  \nn\\
  J_1 = {\sqrt{2}\over s}
  \, M_1^{\mu} \, P_{2\mu}\,, &\quad& J_2 = {\sqrt{2}\over s} \,
  M_2^{\nu} \, P_{1\nu} \,.
  \label{18}
\eea
In the limit $s\to \infty$ the quantity $J_1$ ($J_2$), called impact 
factor, can be calculated, assuming that the energy fractions $x_i$ 
($y_l$) and transverse momenta of the final particles ${\bf p}_{i \perp}$ 
(${\bf p}_{l \perp}$) remain finite. Therefore, the impact factor $J_1$
depends on $x_i,\; {\bf p}_{i\perp}$ with $i\in \;$ jet$_1$ and on the
helicities of the first particle and of the particles in the first jet. Note
that the impact factor $J_1$ ($J_2$) results from the contraction of
the corresponding amplitude with the light--like 4--vector $P_2$
($P_1$).

To show the factorization, we present the metric tensor $g_{\mu
\nu}$ in the form
 \bea
  g^{\mu \nu}&=& \frac{(P_1+P_2)^\mu (P_1+P_2)^\nu }{(P_1+P_2)^2} +
  \nn\\
  &+&
  \frac{(P_1-P_2)^\mu (P_1-P_2)^\nu }{(P_1-P_2)^2} + g_{\perp}^{\mu \nu}=
  \nn \\
  &=&
  \frac{2(P_2^\mu  P_1^\nu+ P_1^\mu P_2^\nu)}{\tilde s} + g_{\perp}^{\mu
  \nu}\,.
  \label{19}
 \eea
The first equality can be easily checked in the cms where
$P_1=(\sqrt{\tilde s}/2) (1,0,0,1)$ and $P_2=(\sqrt{\tilde
s}/2)(1,0,0,-1)$.  Note that Eqs.~(\ref{19}) are exact. Using
this expression for the metric tensor in Eq.~(\ref{17}), we
obtain the amplitude as a sum of three terms
\bea
  M_{fi}&=&\frac{2}{\tilde s q^2}\,\left(M_1^{\mu} P_{2\mu}
  \right)\, \left(M_2^{\nu} P_{1\nu} \right)+
  \nn\\
  &+&
 \frac{2}{\tilde s q^2}\,
  \left(M_1^{\mu} P_{1\mu} \right)\, \left(M_2^{\nu}   P_{2\nu} \right)+
  M_1^{\mu} {g_{\perp\,\mu\nu}\over q^2}   M_2^\nu\,.
  \label{20}
\eea

Let us estimate the contribution of each of these terms to
$M_{fi}$. The amplitude $M_1^{\mu}$ of the upper block depends
on the momentum of the first particle $p_1$, on the momenta
$p_i$ of the particles in the first jet and on the momentum $q$
of the virtual photon. Since $p_1$ and $p_i$ have large
components along $P_1$ and small components along $P_2$ and $q$
has small components both along $P_1$ and along $P_2$, one
obtains the estimates
 \be
  M_1^{\mu} P_{1\mu} \propto s^0\,, \;\; M_1^{\mu} P_{2\mu} \propto s
  \label{21}
\ee
and analogously
\be
  M_2^{\nu} P_{1\nu} \propto s\,, \;\; M_2^{\nu} P_{2\nu} \propto s^0\,.
  \label{22}
 \ee
By virtue of these estimates, only the first term in
Eq.~(\ref{20}) can give a contribution proportional to $s$. As a result, 
we can use the representation (\ref{18}) for the
amplitude of the process.

It is useful to point out another form of Eq.~(\ref{18}). Due to
gauge invariance
of the amplitude with respect to the virtual photon
we have
\be
  q_\mu M_1^\mu = (x_q P_1+y_q P_2 + q_{\perp})_\mu\, M_1^\mu =0\,.
  \label{23}
\ee
Taking into account that $x_q$ is small, one finds
$x_q P_{1\mu}\, M_1^\mu \propto 1/s$,
whereas $y_q P_{2\mu}\,
M_1^\mu \propto s^0$ and $ q_{\perp \mu}\, M_1^\mu \propto s^0$
what leads to 
\be
  M_1^\mu\, P_{2\mu} = - M_1^\mu\, {q_{\perp\,\mu}\over y_q}\,.
  \label{24}
\ee
Analogously, we find for the second amplitude
$ M_2^\mu P_{1\mu}= - M_2^\mu q_{\perp\mu}/x_q$.
Therefore, we can represent the impact factors in the form
\be
  J_1 = -{\sqrt{2}\over sy_q} \,M_1^\mu\, q_{\perp\,\mu}\,,\;\; J_2 =
  -{\sqrt{2}\over sx_q} \,M_2^\nu\, q_{\perp\,\nu}\,.
  \label{25}
 \ee
In other words, up to a factor
$\left[-\sqrt{-2q^2_\perp}/(sy_q)\right]$, the impact factor
$J_1$ coincides with an amplitude describing the scattering of
the first incoming particle with the virtual photon of ``mass''
squared $q^2$ and polarization 4--vector
$q_{\perp\,\mu}/\sqrt{-q^2_\perp}$.

The representations (\ref{25}) of the impact factors are very
important. They show that at small transverse momentum of the
exchanged photon both impact factors should behave as
\be
  J_{1,2} \propto |{\bf q}_\perp| \;\;\mbox{at}\;\;{\bf q}_\perp \to 0\,.
  \label{26}
\ee
In our further analysis we will combine various contributions of
the impact factor into expressions which clearly exhibit such a
behavior. The detailed properties of the impact factors are
described in Sections 4--7.

\subsection{Vertices instead of spinor lines}

Let us consider a virtual electron in the amplitude $M_1$ with
4--momentum $p=(E, {\bf p})$, energy $E>0$ and virtuality
$p^2-m^2$. Due to jet kinematics, its virtuality is small,
$|p^2-m^2| \ll E^2$. We introduce an artificial energy
$$
  E_p = \sqrt{m^2+{\bf p}^2}
$$
and the bispinors  $u_{\bf p}^{(\lambda)}$ and $ v_{\bf
p}^{(\lambda)}$ corresponding to a real electron and a 
real positron
with 3--momentum $ {\bf p}$ and energy $E_p$ (the exact
expressions for these bispinors are given in the Appendix). In
the high-energy limit this artificial energy is close to the
true one:
\be
  \frac{E-E_p}{E+E_p} = \frac{p^2 -m^2}{(E+E_p)^2} \approx
  \frac{p^2 -m^2}{4E^2}\,.
  \label{27}
\ee

Since
\bea
  u_{\bf p}^{(\lambda)} \bar{u}_{\bf p}^{(\lambda)} &=&
  E_p \gamma^0 - {\bf p}\mbox{\boldmath$\gamma$} +m\,,
  \\
  v_{-{\bf p}}^{(\lambda)} \bar{v}_{-{\bf p}}^{(\lambda)} &=&
  E_p \gamma^0 + {\bf p}\mbox{\boldmath$\gamma$} -m\,,
  \nn
\eea
we have the exact identity for the numerator of a virtual
electron \cite{BFKhquasi}:
\be
  \hat p+m = {E+E_p\over 2 E_p} \, u_{\bf p}^{(\lambda)}
  \bar{u}_{\bf p}^{(\lambda)}+ {E-E_p\over 2 E_p} \, v_{-{\bf
  p}}^{(\lambda)} \bar{v}_{-{\bf p}}^{(\lambda)}
  \label{29}
\ee where summation over the helicities $\lambda =\pm 1/2$ is 
understood.
Taking into account the approximation (\ref{27}), we will use
this expression in the simpler form\footnote{
Analogously, for the numerator of the propagator for a virtual
positron with 4--momentum $p=(E, {\bf p})$ and
energy $E>0$ we find the form
$$
  \hat p-m \approx  v_{\bf p}^{(\lambda)} \bar{v}_{\bf p}^{(\lambda)}+ {p^2
  -m^2\over 4 E^2} \, u_{-{\bf p}}^{(\lambda)} \bar{u}_{-{\bf
  p}}^{(\lambda)}\,.
$$ 
}
 \be
  \hat p+m \approx  u_{\bf p}^{(\lambda)}
  \bar{u}_{\bf p}^{(\lambda)}+ {p^2
  -m^2\over 4 E^2} \, v_{-{\bf p}}^{(\lambda)} \bar{v}_{-{\bf
  p}}^{(\lambda)}\,.
  \label{30}
 \ee
Moreover, since
 $$
  v_{-{\bf p}}^{(\lambda)} \bar{v}_{-{\bf p}}^{(\lambda)} =
  E_p \gamma^0 + {\bf p}\mbox{\boldmath$\gamma$} -m =
  x_v {\hat P}_1+y_v {\hat P}_2 + {\hat p}_\perp -m
 $$
with the Sudakov variables (in the given accuracy)
$$
  x_v = {m^2 -p^2_\perp\over 4EE_1}, \;\; y_v = {E\over E_2}\,,
$$
we can present that numerator $\hat p+m$ in another
form
 \be
  \hat p+m \approx u_{\bf p}^{(\lambda)}
  \bar{u}_{\bf p}^{(\lambda)}+ {p^2
  -m^2\over 4 EE_2}\, {\hat P}_2\,.
  \label{31}
 \ee
Omitting terms of the order of (\ref{3}) or higher, the
expressions (\ref{30}), (\ref{31}) are exact.

Using Eq.~(\ref{30}) for all virtual electrons
(of small virtuality) appearing in the impact factors , we are
able to substitute the numerators of all spinor propagators by
transition currents (or generalized vertices) involving real
electrons and real positrons.
As we will show in the next section, those generalized vertices
are finite in the limit $s\to \infty$. On the contrary, a
numerator like $\hat p+m$ is in that limit a sum of a finite
term $\hat p_\perp+m$ and an unpleasant combination $E\gamma^0 -
p_z \gamma_z$ of large terms that requires special care.
Therefore, those replacements significantly  simplify all
calculations of the impact factors $J_i$.

Let us notice some important technical points when calculating
spinor products appearing in the impact factors:

1) If an electron line with numerator $\hat p+m$ connects
vertices with the emission of one real and one virtual photon,
we have the following spinor structure around $\hat p+m$
$$
  {\hat e}^*\, (\hat p+m)\, {\hat P}_2 \;\; {\mbox{ or }} \;\;
  {\hat P}_2\, (\hat p+m)\,{\hat e}^* \,.
$$
In this particular case (using Eq.~(\ref{31}) and taking into
account ${\hat P}_2\,{\hat P}_2\,=0 $) we obtain the simple
substitution rule
\be
  \hat p+m \to  u_{\bf p}^{(\lambda)} \bar{u}_{\bf
  p}^{(\lambda)}\,.
  \label{34}
\ee

{}From that discussion it is obvious that within accuracy
(\ref{3}) the following possible generalized vertices 
\be
  \bar{v}_{-{\bf p}'}^{(\lambda')}\,\hat{P_2}\, v_{-{\bf
  p}}^{(\lambda)}
\label{absent1} 
\ee
will not appear in the calculation of impact
factors.

2) A vertex with an emission of a real photon, ${\hat e}^*$, has
the ``environment'' 
$$
  ({\hat p}' +m)\,{\hat e}^*\,({\hat p} +m)\,.
$$
If we use Eq. (\ref{30}) in the form ${\hat p} +m={\hat a}+
{\hat b}$ and ${\hat p}' +m={\hat a}'+ {\hat b}'$ with ${\hat
a}= u_{\bf p}^{(\lambda)} \bar{u}_{\bf p}^{(\lambda)}$, ${\hat
b}=(p^2-m^2)\, v_{-\bf p}^{(\lambda)} \bar{v}_{-\bf
p}^{(\lambda)}\,/(4E^2)$ and similar expressions for ${\hat
a}'$, ${\hat b}'$, we obtain four terms:
$$
  ({\hat p}' +m)\,{\hat e}^*\,({\hat p} +m) =
  {\hat a}'\, {\hat e}^*\,{\hat a}+
  {\hat a}'\, {\hat e}^*\,{\hat b}+
  {\hat b}'\, {\hat e}^*\,{\hat a}+
  {\hat b}'\, {\hat e}^*\,{\hat b}\,.
$$ 
The last term in this expression is zero, what can be shown as
follows: using for ${\hat b}$ the form ${\hat b}=(p^2-m^2)
\,{\hat P}_2/(4EE_2)$ [see Eq.~(\ref{31})], and a similar one for
${\hat b}'$, and taking into account Eq.~(\ref{14}), one finds
\be
  {\hat b}'\, {\hat e}^*\,{\hat b} \propto
  {\hat P}_2 \,{\hat e}^*\,{\hat P}_2 =
  {\hat P}_2 \,\left( y_{e}{\hat P}_2+{\hat e}_{\perp} \right)^*
  \,{\hat P}_2=0\,.
  \label{35}
\ee
{}From that observation we conclude that in the calculations of
the impact factors generalized vertices of the type
\be
  \bar{v}_{-{\bf p}'}^{(\lambda')}\,\hat{e}^*\, v_{-{\bf
  p}}^{(\lambda)}
 \label{absent2}
\ee
are also absent.

3)
It is easy to check that
$$
  \bar{u}_{\bf p'}^{(\lambda')} \, \hat {e}^{*} \,
  {v}_{-\bf p}^{(\lambda)}\,=
  \bar{v}_{-\bf p'}^{(\lambda')} \, \hat {e}^{*} \,
  {u}_{\bf p}^{(\lambda)}\,.
$$

4) Due to the absence of vertices (\ref{absent1}) and
(\ref{absent2}) generalized vertices of ``exchange'' type can
appear only in pairs (changing the electron state to positron
states with negative 3--momentum and back to an electron state)
in the subsequent emission of two real photons.

\section{Vertices for bremsstrahlung processes}

\subsection{The  $e(p) \to  e(p')+ \gamma(k)$ and
$e(p) + \gamma^*(q) \to e (p')$ transitions}

To calculate the impact factors involving the emission of real
photons we need only two types of vertices: those for the
transition $e(p) \to  e(p')+ \gamma(k)$ where $\gamma(k)$ is a
real photon with helicity $\Lambda$ and the vertex for the
transition $e(p) + \gamma^*(q) \to e (p')$ where $\gamma^*(q)$
is a virtual photon with  energy fraction $x_q=0$ (within our
accuracy).

The following vertices belong to the first type\footnote{We
define the vertices as follows: in writing the amplitude or
impact factor from left to right we follow the electron line
from its beginning to its end. In our case this is more natural
than going  in the opposite direction along the electron line as
usually done.}:
\bea
   V(p,\;k) &\equiv & V_{\lambda \lambda'}^\Lambda(p,\;k) =
  \bar{u}_{\bf p'}^{(\lambda')} \: \hat {e}^{(\Lambda)\,*} \:
  {u}_{\bf p}^{(\lambda)}\,,
  \label{36}\\
  {\tilde V}(p,\;k) &\equiv &
  {\tilde  V}_{\lambda \lambda'}^\Lambda(p,\;k) =
  \bar{u}_{\bf p'}^{(\lambda')} \: \hat {e}^{(\Lambda)\,*} \:
  {v}_{-\bf p}^{(\lambda)}\,=
  \bar{v}_{-\bf p'}^{(\lambda')} \: \hat {e}^{(\Lambda)\,*} \:
  {u}_{\bf p}^{(\lambda)}\,.
  \nn\\
  \label{37}
\eea
Certainly, the calculation of these vertices does not
depend on the concrete representation of bispinors and
$\gamma$-matrices, but it is very convenient to use the spinor
representation described in the Appendix. The result of
calculation with accuracy (\ref{3}) is the following
\bea
  V(p,\;k)&=&\left[\delta_{\lambda \lambda'}\,
  2\, \left( {e}^{(\Lambda)\,*} p \right)\, \left(1- x\,
  \delta_{\Lambda,- 2\lambda}\right) + 
  \right.
  \nn\\ 
  &+& \left. \delta_{\lambda,-
  \lambda'}\,\delta_{\Lambda, 2\lambda}\, \sqrt{2}\, mx
  \,\right]\, \Phi\,,
  \label{38}\\
  {\tilde V}(p,\;k) &=& -2\sqrt{2}\, \Lambda\, E'\,
  \delta_{\lambda,- \lambda'}\,\delta_{\Lambda, 2\lambda}\,\Phi
  \label{39}
\eea
where
\be
  x={\omega\over E}\,,\;\;\; \Phi = \sqrt{E\over E'}\, {\mathrm
  e}^{{\rm i} (\lambda' \varphi' - \lambda \varphi)}\,.
  \label{40}
\ee
It is useful to remind here that for the polarization vectors
$e$ [see Eq.~(\ref{14})] we have
\be
{e}p = {e}_\perp\, \left(p_\perp - {k_\perp \over x} \right)\,.
 \label{41}
\ee

The vertex  of the second type is very simple:
\be
  V(p) \equiv {V}_{\lambda \lambda'} (p) ={\sqrt{2}\over s}\;
  \bar{u}_{\bf p'}^{(\lambda')}\,  \hat {P}_2 \,{u}_{\bf
  p}^{(\lambda)} = \sqrt{2}\, {E\over E_1}\, \delta_{\lambda
  \lambda'}\, \Phi\,.
  \label{42}
\ee

Let us make some general remarks related to
Eqs.~(\ref{38})--(\ref{42}):

1)
 We have previously mentioned [after Eq.~(\ref{18})] that the
impact factors remain finite in the high--energy limit $s\to
\infty$. Now we observe from (\ref{38}), (\ref{41}) and
(\ref{42}) that the vertices $V(p,k)$ and $V(p)$ are finite in
that limit too. Due to properties 3) and 4) discussed in Sect.
2.3 the ``exchange'' vertices $\tilde V$ appear only in
combinations
$[(p^2-m^2)/(4E^2)] {\tilde V}(p-k_i, k_i){\tilde V}(p,k_{i+1})$
which also remain finite. Indeed, the factor $p^2-m^2$
(denominator of the considered lepton propagator between
neighbouring real photons) gives the finite virtuality in the
high energy limit, the $1/E^2$ factor  combines with the
energies $E$ and $E'=E-\omega_{i+1}$ according to Eq.~(\ref{37})
to an energy independent factor. All calculations are exact up
to neglected pieces of the order of (\ref{3}).

2)
For the bremsstrahlung of $n$ real photons along
an electron line, the production of factors $\Phi$,
corresponding to the emission of $n$ real photons and one
virtual photon,
\be
  \Phi_1 \Phi_2... \Phi_{n} \Phi_q = \sqrt{E_1\over E_3}\,
  {\mathrm e}^{{\rm i}(\lambda_3 \varphi_3 -\lambda_1 \varphi_1)}
  \label{43}
\ee
is proportional to a phase factor which can be included in the
definition of the corresponding amplitude. Therefore, in the
calculation of this amplitude we can omit all factors $\Phi$
appearing in Eqs. (\ref{38}), (\ref{39}), (\ref{42}).

 3)
Up to now we have considered the bremsstrahlung by electrons. It
is quite natural that the presented formulae are also valid for
the bremsstrahlung by positrons. Let us consider, for example,
the vertex $\bar{v}_{\bf p}^{(\lambda)} \, \hat
{e}^{(\Lambda)\,*} \, {v}_{\bf p'}^{(\lambda')}$ which
corresponds to the $e^+(p) \to e^+(p')+ \gamma(k)$ transition.
If we take into account the relations for bispinors
\bea
  {v}_{\bf p'}^{(\lambda')}&=&
    C\,\left(\bar{u}_{\mathbf p'}^{(\lambda')}\right)^T\,,\;\;\;
   \\
      \bar{v}_{\mathbf p}^{(\lambda)}&=& \left( C\,{u}_{\bf
   p}^{(\lambda)}\right)^T= \left({u}_{\bf p}^{(\lambda)}
   \right)^T \,C^{T}= -\left({u}_{\bf p}^{(\lambda)}
   \right)^T \,C^{-1} \nn
 \label{43a}
\eea
and $\gamma$ matrices
  \be
    C^{-1} \hat{e}^{(\Lambda)\,*}C =
    - \left(\hat{e}^{(\Lambda)\,*}\right)^T
    \label{43b}
     \ee
(here the matrix $C=\gamma^2\gamma^0$ is related to the charge
conjugation operator, see Appendix), we immediately obtain
\be
  \bar{v}_{\bf p}^{(\lambda)} \, \hat {e}^{(\Lambda)\,*} \,
  {v}_{\bf p'}^{(\lambda')}=\bar{u}_{\bf p'}^{(\lambda')} \, \hat
  {e}^{(\Lambda)\,*} \, {u}_{\bf p}^{(\lambda)}\,\equiv\,
  V(p,\;k)\,.
\ee

\subsection{Helicity conserved and helicity non--conserved vertices}

In the further calculations we need only  formulae
(\ref{36})--(\ref{43}). But for reference reasons it is
convenient to rewrite them for some particular cases, omitting
the factors $\Phi$.

In the case of the helicity conserved (HC) transitions,
$\lambda' = \lambda$, the vertices are of the form:
 \bea
  V(p,k)& =& 2e^* p \;\;\;\;\;\;\;\;\;\; \;\;\;\,
  \mbox{ for } \;\;
  \Lambda =2\lambda =2 \lambda'\,,
  \label{44}\\
  V(p,k) &=& 2e^* p\, (1-x) \;\; \mbox{ for } \;\; \Lambda
  =-2\lambda =-2 \lambda'\,,
  \label{45}\\
  V(p) &=& \sqrt{2} \, E/E_1\,,
  \label{46}\\
  \tilde V(p,k) &=& 0
  \label{47}
\eea
and additionally [taking into account Eq.~(\ref{41})]
\bea
  &&2e^{(+)*} p = \sqrt{2}\, z^*\,,\;\;
  2e^{(-)*} p = -\sqrt{2}\, z\,, \;\;
  \nn\\
  &&z = p_x + {\rm i} p_y - { k_x + {\rm i} k_y\over x}\,.
  \label{48}
\eea

In the case of the helicity non--conserved (HNC) transitions,
$\lambda' =- \lambda$, we have:\footnote{Here and in the
following we use the sign notation both for the photon
polarization $\Lambda=\pm 1=\pm$ and the lepton helicity
$\lambda=\pm 1/2=\pm\,$.}
 \bea
  V_{+-}^+(p,k) &=&
  V_{-+}^-(p,k) = \sqrt{2}\, mx\,,
  \label{49}\\
  \tilde V_{+-}^+(p,k) &=& - \tilde V_{-+}^-(p,k) =
  -2\sqrt{2}\, E'\,,
  \label{50}\\
  V_{+-}^-(p,k) &=& V_{-+}^+(p,k) = \tilde V_{+-}^-(p,k)=
  \nn\\
  &=& \tilde V_{-+}^+(p,k)= V_{+-}(p)= V_{-+}(p) =0\,.
  \label{51}
 \eea

\subsection{Properties of vertices}

{}From the derived expressions for the vertices the following
properties can be found:

1) Vertices with a maximal change of helicity,
\be
  \max |\Delta \lambda|= \max|\Lambda+ \lambda' -\lambda|=2\,,
  \label{52}
 \ee
are absent what can be seen from Eqs.~(\ref{51}). This
property as well as property 4) below are a result of the
conservation of the total angular momentum $J_z$ in the strict
forward direction. Indeed, the vertices for HNC transitions do
not depend on the transverse momenta of the particles in the
jet. Therefore, they do not change for transitions setting all
transverse momenta to zero. In other words, those vertices can
be calculated for the case of strict forward emission for which
the total angular momentum is conserved: $J_z = \lambda =
\Lambda+\lambda'$.

2) If the produced photon becomes very hard ($\omega \to E$) the
initial electron ``transmits'' its helicity to the photon:
 \be
  V (p,k) \propto  \delta_{\Lambda, 2\lambda}\,,\;\;\;
 \tilde V (p,k) \to 0\;\;\; \mbox{ for } \;\; x\to 1 \,.
  \label{53}
 \ee

3) If the final electron becomes hard ($E' \to E$, soft photon
limit, $x\ll1$), the initial electron ``transmits'' its helicity
to the final electron: in that limit the vertex
 \be
   V(p,k) = - {2\over x}\left( e_\perp^*\, k_\perp \right)
   \,  \delta_{\lambda \lambda'}
  \label{54}
 \ee
dominates, which corresponds to the approximation of a classical
current.

4) For HNC vertices a strong correlation between the
helicities of the initial electron and the photon exists:
\be
  \Lambda = 2 \lambda \; \;\;\mbox{ if }\;\;\;
  \lambda'=-\lambda\,.
  \label{55}
\ee For HC vertices there is no strong correlation between
helicities of electrons and the photon (excluding the limiting
case of $\omega \to E$).

5) From Eqs.~(\ref{44}), (\ref{45}), (\ref{48}) it can be seen
that
\be
  V_{\lambda \lambda}^{+} \propto \, z^*\,,\;\;
  V_{\lambda \lambda}^{-} \propto \, z
 \label{56}
\ee
where $z$ is defined in Eq.~(\ref{48}).

\section{Impact factor for the single bremsstrahlung 
$e(p_1)+ \gamma^*(q) \to e(p_3)+ \gamma(k)$ }

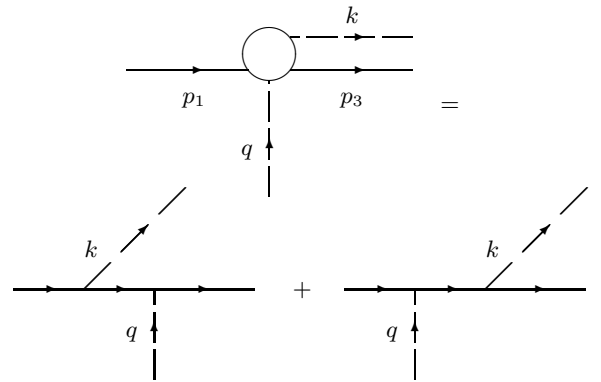
\begin{figure}[!htb]
  \begin{center}
  \unitlength=2.00mm
  \begin{picture}(25.00,12.00)(2.50,2.50)
     \put(12.00,12.00){\circle{3.40}}
     \put( 2.50,10.90){\line(1,0){8.10}}
     \put( 2.50,10.90){\vector(1,0){5.15}}
     \put(12.00, 2.50){\line(0,1){2.0}}
     \put(12.00, 5.00){\line(0,1){2.0}}
     \put(12.00, 5.00){\vector(0,1){1.40}}
     \put(12.00, 7.50){\line(0,1){2.0}}
     \put(12.00,10.00){\line(0,1){0.2}}
     \put(13.40,10.90){\line(1,0){ 8.1}}
     \put(13.40,10.90){\vector(1,0) {5.00}}
     \put(13.40,13.10){\line(1,0){0.6}}
     \put(14.50,13.10){\line(1,0){2.0}}
     \put(17.00,13.10){\line(1,0){2.0}}
     \put(19.50,13.10){\line(1,0){2.0}}
     \put(17.00,13.10){\vector(1,0){1.40}}
     \put(17.50,14.60){\makebox(0,0)[cc]{$k$}}
     \put(17.50, 8.90){\makebox(0,0)[cc]{$p_3$}}
     \put( 7.00, 8.90){\makebox(0,0)[cc]{$p_1$}}
     \put(10.50, 5.50){\makebox(0,0)[cc]{$q$}}
     \put(24.00, 8.50){\makebox(0,0)[cc]{$=$}}
   \end{picture}
   \begin{picture}(61.00,12.00)(25.0,2.50)
     \put(49.00, 8.50){\line(1,0){9.0}}
     \put(49.00,8.50){\vector(1,0){2.90}}
     \put(53.66,8.50){\vector(1,0){2.90}}
     \put(56.00,8.50){\line(1,0){9.0}}
     \put(61.40,8.50){\vector(1,0){1.0}}
     \put(53.66, 2.50){\line(0,1){2.00}}
     \put(53.66,5.00){\line(0,1){2.00}}
     \put(53.66,7.50){\line(0,1){1.00}}
     \put(53.66,5.00){\vector(0,1){1.40}}
     \put(63.34,13.50){\line(1,1){1.8}}
     \put(60.84,11.00){\line(1,1){1.8}}
     \put(58.34,8.50){\line(1,1){1.8}}
     \put(60.84,11.00){\vector(1,1){1.8}}
     \put(58.84,11.20){\makebox(0,0)[cc]{$k$}}
     \put(52.16,5.50){\makebox(0,0)[cc]{$q$}}
     \put(46.20, 8.50){\makebox(0,0)[cc]{$+$}}
     \put(27.00, 8.50){\line(1,0){ 9,0}}
     \put(27.00,8.50){\vector(1,0){2.90}}
     \put(31.66,8.50){\vector(1,0){2.90}}
     \put(34.00, 8.50){\line(1,0){ 9.0}}
      \put(34.00,8.50){\vector(1,0){6.00}}
     \put(32.16,11.20){\makebox(0,0)[cc]{$k$}}
     \put(34.84,5.50){\makebox(0,0)[cc]{$q$}}
     \put(36.34, 2.50){\line(0,1){2.00}}
     \put(36.34,5.00){\line(0,1){2.00}}
     \put(36.34,7.50){\line(0,1){1.00}}
     \put(36.34, 5.00){\vector(0,1){1.4}}
     \put(36.66,13.50){\line(1,1){1.80}}
     \put(34.16,11.00){\line(1,1){1.80}}
     \put(31.66,8.50){\line(1,1){1.80}}
     \put(34.16,11.00){\vector(1,1){1.8}}
  \end{picture}
  \end{center}
  \caption{Amplitude for the virtual Compton scattering.}
  \label{fig:12}
\end{figure}

The impact factor for the single bremsstrahlung corresponds  to
the virtual Compton scattering (Fig.~\ref{fig:12})
where $J_1$ is given as follows:
\be
  J_1 (e_{\lambda_1}+\gamma^* \to e_{\lambda_3}+ \gamma_\Lambda )=
  4\pi\alpha\, \left(\frac{N_1}{2p_1k} - \frac{N_3}{2p_3k}\right)
  \label{57}
\ee
with
\bea
  N_1&=& \bar u_3 \frac{\sqrt{2} {\hat P}_2}{s}\,
  (\hat{p}_1-\hat k +m)\hat{e}^* u_1\,,\;\;
  \nn\\
  N_3&=& \bar u_3 \hat{e}^* (\hat{p}_3 +\hat k +m)
   \frac{\sqrt{2} {\hat P}_2}{s} \, u_1\,,
  \label{58}
\eea
$e \equiv e^{(\Lambda)}(k)$ is the polarization 4-vector of the
final photon. Here for $\hat{p}_1-\hat k +m$ and $\hat{p}_3
+\hat k +m$ we can use the simple substitution (\ref{34}) that
allows us to eliminate the numerators of the two spinor
propagators and to introduce the vertices $V(p)$ and $V(p,k)$.

The vertices $V(p)$ are diagonal in the helicity basis and
simply lead to factors $\sqrt{2}\,(1-x)$ with $x=\omega/E_1$ for
$N_1$ and $\sqrt{2}$ for $N_3$. As a result, we have
 \be
  J_1 = \sqrt{2}\, 4\pi \alpha\,  \left[ {1-x\over 2p_1k} V(p_1,k) -
  {1\over 2p_3k}\,V(p_3+k,k) \right]\,\Phi\,\,,\;\;
  \label{59}
\ee
with the vertices
(remind that $ e^{(\Lambda)\,*}\,k=0$)
\bea
  V(p_1,\;k) &=& \delta_{\lambda_1 \lambda_3}\, 2\,
  \left(  {e}^{(\Lambda)\,*}
  p_1 \right)\, \left(1- x\, \delta_{\Lambda,-
  2\lambda_1}\right) +
  \nn\\
  &+&
  \delta_{\lambda_1,- \lambda_3}\,\delta_{\Lambda, 2\lambda_1}\,
  \sqrt{2}\, m x\,,
  \label{60}\\
  V(p_3+k,\;k) &=& \delta_{\lambda_1 \lambda_3}\, 2\, \left(
  {e}^{(\Lambda)\,*} p_3 \right)\, \left(1- x\,
  \delta_{\Lambda,-
  2\lambda_1}\right) + 
  \nn\\
  &+&
  \delta_{\lambda_1,- \lambda_3}\,
  \delta_{\Lambda,
  2\lambda_1}\, \sqrt{2}\, m x
  \nn
 \eea
and the $\Phi$ factor in the form of Eq.~(\ref{43})
 \be
  \Phi = {1\over \sqrt{1-x}} \, {\mathrm e}^{{\rm i}(\lambda_3
  \varphi_3 -\lambda_1 \varphi_1)}\,.
  \label{61}
 \ee

{}From Eq.~(\ref{59}) it is clear that the properties of $J_1$
are determined by the properties of the vertices described in
Sect.~3.3. In the soft photon limit, $x\ll 1$, we have the usual
approximation by classical currents:
 \be
  J_1 = \sqrt{2}\, 4\pi \alpha\,  \left( {e^*p_1\over p_1k} -
  {e^*p_3\over p_3k}\right) \, \Phi\:\delta_{\lambda_1 \lambda_3}\,.
  \label{62}
 \ee

The impact factor $J_1$ can be transformed to a form which
clearly exhibits the proportionality $J_1 \propto q_\perp$
resulting from the gauge invariance of $J_1$ with respect to the
virtual photon (see Eq. (\ref{26})). For this purpose we use
Eqs.~(\ref{60}), (\ref{41}) and rewrite
\bea
  V(p_3+k,k)&=& V(p_1 +q,k) = V(p_1,k) + 
  \\&+& 2 \,
  \left(q_\perp\,{e}^{(\Lambda)\,*}_\perp \right)\, \left( 1- x\,
  \delta_{\Lambda,- 2\lambda_1} \right)\, \delta_{\lambda_1
  \lambda_3}\,.\nn
  \label{63}
\eea
This gives the following result
\bea
  J_1 &=& \sqrt{2}\, 4\pi \alpha\,  \left[ A_1\,V(p_1,k)+q_\perp
  B_1\right]\,\Phi\,,
  \label{64} \\
  A_1&=&{1-x\over 2p_1k} -{1\over 2p_3k}\,,\;
  B_1 =-{{e}^{(\Lambda)\,*}_\perp \over p_3k}\, \left( 1- x\,
  \delta_{\Lambda,- 2\lambda_1} \right)\, \delta_{\lambda_1
  \lambda_3} \,.
  \nn
 \eea
The last term in $J_1$  is directly proportional to $q_\perp$
and it is not difficult to check that the same is true for the
first term. Indeed, since
\begin{eqnarray*}
  2p_1k &=& x a\,, \;\; a=\,m^2 + {{\bf k}^2_\perp \over x^2}\,,
  \nn\\
  2p_3k &=& {x\over 1-x}\,b\,, \;\; b=m^2 +
  \left({\bf q}_\perp
  - {{\bf k}_\perp \over x}\right)^2\,,
\end{eqnarray*}
we immediately obtain
\be
  A_1={1-x\over x}\,\left( {1\over a} - {1\over b}\right)
  \propto  q_\perp \,.
  \label{65}
 \ee
As a result, Eq. (\ref{64}) is a simple and compact expression
for all 8 helicity states written in such a form that all
individual large (compared to $ q_\perp$) contributions are
cancelled.

Let us discuss the form of $J_1$ for the single bremsstrahlung by
a positron. We expect that the only difference is connected with
the change of the charge sign $-e \to +e$ in each vertex with
the emission of a real or virtual photon. In our case this gives
the additional factor $(-1)^2 =1$, therefore,
 \be
  J_1 \left(e^+_{\lambda_1}+\gamma^* \to e^+_{\lambda_3}+
  \gamma_{\Lambda }\right)=
  J_1 \left(e^-_{\lambda_1}+\gamma^* \to e^-_{\lambda_3}+
  \gamma_{\Lambda }\right)\,.
 \label{65a}
 \ee

To give the formal proof of this relation, we take into account
that going over from electron to positron bremsstrahlung the
numerators of the electron propagators $\hat p +m$ have to be
replaced by $-\hat p+m$ for the positrons and the bispinors
$u_1$ and $\bar{u}_3$   for the electrons by those for the
positrons $\bar{v}_1$ and $v_3$. In addition , a factor $(-1)$
has to be added according to one of the Feynman
rules\footnote{See, for example, rule (9) in
text-book~\cite{BLP} \S 77: "An additional factor $-1$ is
included in ${\rm i} M_{fi}$ for ... each pair of positron
external lines if these are beginning and end of a single
sequence of a lepton line."}. It gives
 $$
  J_1 \left(e^+_{\lambda_1}+\gamma^* \to e^+_{\lambda_3}+
  \gamma_\Lambda  \right)= (-1)\cdot
  4\pi\alpha\, \left(\frac{\tilde{N}_1}{2p_1k} -
  \frac{\tilde{N}_3}{2p_3k}\right)
  \label{57a}
 $$
with
\begin{eqnarray*}
  \tilde{N}_1&=& \bar{v}_1\,\hat{e}^*
  (-\hat{p}_1+\hat k +m) \frac{\sqrt{2} {\hat P}_2}{s}\, v_3\,,
   \nn\\
  \tilde{N}_3&=& \bar{v}_1\, \frac{\sqrt{2} {\hat P}_2}{s} \,
  (-\hat{p}_3 -\hat k +m)\,\hat{e}^*  v_3\,.
\end{eqnarray*}
If we take into account (see Appendix and Eqs. (\ref{43a}),
(\ref{43b})) the relations for bispinors
\bea
  &&(-1)\cdot \bar{v}_{1}=-\left(C\,{u}_{1}\right)^T
  =\left({u}_{1}\right)^T C^{-1}\,,
  \nn\\
  && {v}_{3}=  C\,\left(\bar{u}_{3}\right)^T\,,
  \label{65b}
\eea
spinor propagators
\be
  C^{-1}(-\hat p +m)C = (\hat p +m)^T
  \label{65c}
\ee
and vertices for the real and virtual photons
\be
  C^{-1} \hat{e}^*C = - \left(\hat{e}^*\right)^T\,,\quad
  C^{-1} \hat{P}_2 C = - \left(\hat{P}_2\right)^T
  \label{65d}
\ee
with the matrix $C=\gamma^2\gamma^0$, we immediately obtain 
Eq.~(\ref{65a}).

The basic Eq. (\ref{59}) can be rewritten in another form which
may be useful in concrete calculations. If we take into account
that
 \be
 e^*p_1 = -{e^*_\perp k_\perp\over x}\,,\;\; e^*p_3 = e^*(p_1+q) =
e^*_\perp\left(q_\perp- {k_\perp\over x}\right)\,,
 \label{66}
 \ee
we arrive at the result of Ref.~\cite{kuraevzp86}:
\bea
&&J_1 (e_{\lambda_1}+\gamma^* \to e_{\lambda_3}+ \gamma_{\Lambda
})= 
\nn\\
 &&=8\pi\alpha {\sqrt{1-x}\over x}\, {\mathrm e}^{{\rm i}
(\lambda_3 \varphi_3 - \lambda_1 \varphi_1)} \times
 \label{67} \\
 &&\left[ \left(1- x \delta_{\Lambda,- 2\lambda_1}\right)\,
 \sqrt{2} {\bf T} {\bf e}^{(\Lambda)\,*}_\perp \,
 \delta_{\lambda_1 \lambda_3}+
 m \,x S \delta_{\lambda_1, -\lambda_3}
 \delta_{\Lambda, 2 \lambda_1} \right]
 \nn
 \eea
where the transverse 4--vector $T$ (in the used reference frame
$T=(0,\mathbf T,0), \, \, T^2=-{\mathbf T}^2$) and the scalar
$S$ are defined as
\begin{equation}
T={(k_\perp/x) \over a} + {q_\perp -(k_\perp/x)\over b}\,,
\;\;\; S={1\over a} - {1\over b}
 \label{68}
 \end{equation}
with the useful relation
\begin{equation}
{\bf T}^2 + m^2 S^2 = {{\bf q}_\perp^2 \over ab}\,.
 \label{69}
 \end{equation}
Since
 \be
T \propto q_\perp\,,  \;\;\; S\propto q_\perp\,,
 \label{70}
 \ee
we again conclude that $J_1 \propto q_\perp$.

\section{\hspace{-1.5mm} Impact factor for the double bremsstrahlung 
$e(p_1)+ \gamma^*(q) \to e(p_3)+ \gamma(k_1)+ \gamma(k_2)$}

\subsection{Notations}

\begin{figure}[!htb]
  \begin{center}
  \unitlength=2.00mm
  \begin{picture}(45.00,16.00)(0.00,0.00)
    \put(1.00, 8.50){\line(1,0){9.0}}
    \put(1.00,8.50){\vector(1,0){1.50}}
    \put(4.66,8.50){\vector(1,0){2.90}}
    \put(7.00,8.50){\vector(1,0){11.0}}
    \put(12.40,8.50){\vector(1,0){1.0}}
    \put(2.00,7.00){\makebox(0,0)[cc]{$p_1$}}
    \put(17.00,7.00){\makebox(0,0)[cc]{$p_3$}}
    \put(14.00, 2.50){\line(0,1){2.00}}
    \put(14.00,5.00){\line(0,1){2.00}}
    \put(14.00,7.50){\line(0,1){1.00}}
    \put(14.00,5.00){\vector(0,1){1.40}}
    \put(12.00,5.50){\makebox(0,0)[cc]{$q$}}
    \put(14.00,13.50){\line(1,1){1.8}}
    \put(11.50,11.00){\line(1,1){1.8}}
    \put(9.00,8.50){\line(1,1){1.8}}
    \put(11.50,11.00){\vector(1,1){1.8}}
    \put(9.00,11.20){\makebox(0,0)[cc]{$k_2$}}
    \put(9.00,13.50){\line(1,1){1.8}}
    \put(6.50,11.00){\line(1,1){1.8}}
    \put(4.00,8.50){\line(1,1){1.8}}
    \put(6.50,11.00){\vector(1,1){1.8}}
    \put(4.00,11.20){\makebox(0,0)[cc]{$k_1$}}
    \put(23.00, 8.50){\line(1,0){9.0}}
    \put(23.00,8.50){\vector(1,0){1.50}}
    \put(26.66,8.50){\vector(1,0){2.90}}
    \put(29.00,8.50){\vector(1,0){11.0}}
    \put(34.40,8.50){\vector(1,0){1.0}}
    \put(31.00, 2.50){\line(0,1){2.00}}
    \put(31.00,5.00){\line(0,1){2.00}}
    \put(31.00,7.50){\line(0,1){1.00}}
    \put(31.00,5.00){\vector(0,1){1.40}}
    \put(29.00,5.50){\makebox(0,0)[cc]{$q$}}
    \put(31.00,13.50){\line(1,1){1.8}}
    \put(28.50,11.00){\line(1,1){1.8}}
    \put(26.00,8.50){\line(1,1){1.8}}
    \put(28.50,11.00){\vector(1,1){1.8}}
    \put(26.00,11.20){\makebox(0,0)[cc]{$k_1$}}
    \put(41.00,13.50){\line(1,1){1.8}}
    \put(38.50,11.00){\line(1,1){1.8}}
    \put(36.00,8.50){\line(1,1){1.8}}
    \put(38.50,11.00){\vector(1,1){1.8}}
    \put(36.00,11.20){\makebox(0,0)[cc]{$k_2$}}
  \end{picture} 
  \begin{picture}(65.00,16.00)(35.00,0.00)
    \put(45.00, 8.50){\line(1,0){9.0}}
    \put(45.00,8.50){\vector(1,0){1.50}}
    \put(48.66,8.50){\vector(1,0){2.90}}
    \put(51.00,8.50){\vector(1,0){11.0}}
    \put(56.40,8.50){\vector(1,0){1.0}}
    \put(48.00, 2.50){\line(0,1){2.00}}
    \put(48.00,5.00){\line(0,1){2.00}}
    \put(48.00,7.50){\line(0,1){1.00}}
    \put(48.00,5.00){\vector(0,1){1.40}}
    \put(46.00,5.50){\makebox(0,0)[cc]{$q$}}
    \put(58.00,13.50){\line(1,1){1.8}}
    \put(55.50,11.00){\line(1,1){1.8}}
    \put(53.00,8.50){\line(1,1){1.8}}
    \put(55.50,11.00){\vector(1,1){1.8}}
    \put(53.00,11.20){\makebox(0,0)[cc]{$k_1$}}
    \put(63.00,13.50){\line(1,1){1.8}}
    \put(60.50,11.00){\line(1,1){1.8}}
    \put(58.00,8.50){\line(1,1){1.8}}
    \put(60.50,11.00){\vector(1,1){1.8}}
    \put(58.00,11.20){\makebox(0,0)[cc]{$k_2$}}
  \end{picture}
  \end{center}
  \caption{Feynman diagrams for the impact factor related to the
  double bremsstrahlung, diagrams with $k_1 \leftrightarrow k_2$
  photon exchange have to be added.}
  \label{fig:13}
\end{figure}
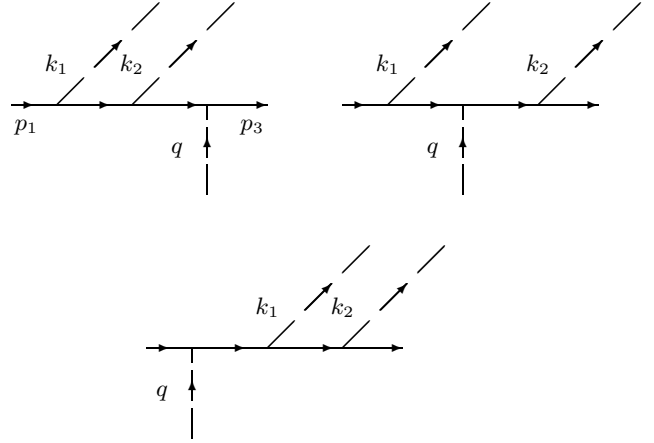

The impact factor for the double bremsstrahlung corresponds to
six diagrams, three of them are shown in Fig.~\ref{fig:13}.
We indicate explicitly the helicity states of the initial and
final electrons $\lambda_{1,3}$ and of the final photons
$\Lambda_{1,2}$\footnote{Here the
amplitude ${\cal M}$ differs from the same quantity used
in Ref.~\cite{KSSSh00} by a factor $S_{\lambda_{e3}}$.}:
\be
   J_{1}= \sqrt{2} \, (4\pi\alpha)^{3/2}\, X_3 \,
   {\cal M}_{\lambda_1\, \lambda_3}^{\Lambda_1\, \Lambda_2}\, (x_1, x_2,
   k_{1\perp},  k_{2\perp}, p_{3\perp}) \, \Phi 
   \label{71} 
\ee
with
\be
   \Phi= \frac{1}{\sqrt{X_3}} \,
   {\mathrm e}^{{\rm i} (\lambda_3 \varphi_3 - \lambda_1 \varphi_1)} \,.
   \label{71a}
\ee
Let us stress again that $J_{1}$ and ${\cal M}$ do not depend on
$s$. They depend only on the energy fractions
 $$
x_{1,2} =\omega_{1,2}/E_1, \;\; X_3 = E_3/E_1\,,\;\;\;
 x_1+x_2+X_3=1
 $$
and on the transverse momenta of the final particles in the
first jet.

We also introduce the transverse vectors ($j=1,\,2$)
\be
  q_\perp = k_{1\perp} +  k_{2\perp} +p_{3\perp}\,, \quad
  r_j= (X_3k_j-x_j p_3)_\perp\,, 
  \label{72}
 \ee
and useful complex combinations of the transverse vector
components [cf. Eq.~(\ref{48})]
 \be
  K_j = k_{jx} + {\rm i}\, k_{jy}, \;\; Q = q_{x} + {\rm i}\, q_{y},
  \;\; R_j = r_{jx} + {\rm i}\, r_{jy}\,.
  \label{73}
\ee
The polarization 4-vectors $e_j\equiv e^{(\Lambda_j)}(k_j)$
for both final photons are chosen in the
form~(\ref{14})--(\ref{16}).

The denominators of the propagators in Fig.~\ref{fig:13} are
expressed via the energy fractions and transverse momenta as
follows:
\bea
  \label{75}
   a_j&\equiv&-(p_1-k_j)^2+m^2=
   \frac{1}{x_j}(m^2x_i^2+{\mathbf k}_{j\perp}^2) \,,
  \nonumber
  \\
  b_j&\equiv&(p_3+k_j)^2-m^2=\frac{1}{x_j X_3}(m^2 x_j^2+
  {\mathbf r}_{j\perp}^2) \,,
  \nonumber
  \\
  a_{12}=a_{21}&\equiv&-(p_1-k_1-k_2)^2+m^2=
  \nn\\
  &=&a_1+ a_2 -
  \frac{1}{x_1x_2}
  \left(x_1{\mathbf{k}}_{2\perp}
  -x_2{\mathbf{k}}_{1\perp} \right)^2\,,
\nonumber
\\
  b_{12}=b_{21}&\equiv&(p_3+k_1+k_2)^2-m^2=
  \nn\\
   &=&b_1+
  b_2+\frac{1}{x_1x_2}
  \left(x_1{\mathbf{k}}_{2\perp}
  -x_2{\mathbf{k}}_{1\perp} \right)^2\,.
\eea

\subsection{General formula for the helicity amplitudes}

Following the electron line from left to right in the  Feynman
diagrams in Fig.~\ref{fig:13} and writing down the corresponding
vertices, we immediately obtain the result for the impact factor
$J_1$ or the amplitude ${\cal M}$ of Eq.~(\ref{71}). Moreover,
if the electron line begins or ends at a vertex with the virtual
photon (the first and last diagram of Fig.~\ref{fig:13}), we can
use the simple substitution (\ref{34}) just as in Sect 4. In the
other case we use the substitution rule~(\ref{30}).
Therefore, for the first and the third diagrams we get
contributions as products of two adjacent vertices with real
photon emission from electrons
with the simple vertex including the
virtual photon. In the case of the second diagram we have only
the product of  two vertices with real photons emission from the
electron and the    $e\gamma^*\to e$ transition vertex located
in between those vertices.

Taking into account that the $e\gamma^*\to e$ transition
vertices $V(p)$ lead to energy fraction factors
$\sqrt{2}\,X_3$, $\sqrt{2}\,(1-x_1)$ and $\sqrt{2}$ for the
first, second and third diagram  of Fig. \ref{fig:13}, respectively, we
find\footnote{Let us remind that $M$, $V$ and $\tilde V$ are
matrices
with respect to lepton helicities, in particular,
 $$
  M= M_{\lambda_1\, \lambda_3}^{\Lambda_1\, \Lambda_2}\,, \;\;
  V(p,k_1) = V_{\lambda_1\, \lambda}^{\Lambda_1} (p, k_1)\,,\;\;
  V(p,k_2) = V_{\lambda\, \lambda_3}^{\Lambda_2} (p, k_2)
 $$
and that in Eq.~(\ref{77}) the summation over $\lambda$ is
assumed.}
 \be
  {\cal M}= \left(1 +{\cal P}_{12}\right)\, M,
  \label{76}
 \ee
 \bea
  X_3 M &=& \frac{X_3}{a_1 a_{12}}\, V(p_1, k_1)\,
  V(p_1-k_1, k_2) - 
  \nn\\
  &-&
  \frac{1-x_1}{a_1 b_2}\, V(p_1, k_1)\, V(p_1-k_1+q, k_2)+
  \nn\\
  &+&
  \frac{1}{b_{12} b_2}\, V(p_1 +q, k_1)\, V(p_1-k_1+q, k_2)-
  \nn\\
  &-&
  \frac{X_3}{a_{12}}\, \frac{\tilde V(p_1, k_1)\, \tilde
  V(p_1-k_1, k_2)} {4(E_1-\omega_1)^2} +
  \nn\\
  &+&
  {1\over b_{12}}\,{\tilde V(p_1 +q, k_1)\,
  \tilde V(p_1-k_1+q, k_2)
  \over 4(E_3+\omega_2)^2}\,.
  \label{77}
  \eea
The permutation operator ${\cal{P}}_{12}$ for the photons
is defined as
\begin{eqnarray*}
  {\cal{P}}_{12}f(k_1,e_1;k_2,e_2)= f(k_2,e_2;k_1,e_1)\,, \qquad
  {\cal{P}}_{12}^2 =1.
\end{eqnarray*}

Now we transform $J_1$ to a form which clearly exhibits the
proportionality $J_1 \propto q_\perp$. Using the following
properties of vertices
\bea
  V(p_1 +q,k_1) &=& V(p_1,k_1) +
  \nn\\
  &+&
  2\left({e}^{(\Lambda_1)\,*}_\perp\, q_\perp \right) \left( 1-
  x_1\, \delta_{\Lambda_1,- 2\lambda_1} \right)\,
  \delta_{\lambda_1 \lambda} \,,
  \nn\\
  V(p_1-k_1 , k_2)& = &V(p_1-k_1 +q, k_2) -
  2\left({e}^{(\Lambda_2)\,*}_\perp\, q_\perp \right) 
  \nn\\
  &\times&
  \left( 1-
  {x_2\over 1-x_1}\, \delta_{\Lambda_2,- 2\lambda} \right)\,
  \delta_{\lambda \lambda_3}\,,
  \label{78}\\
  \tilde V(p_1 +q, k_1)&=& \tilde V(p_1, k_1)\,,
  \nn\\
  \tilde
  V(p_1-k_1 +q, k_2)&=& \tilde V(p_1-k_1, k_2)\,.
   \nn
 \eea
we obtain the result\footnote{We indicate explicitly all
external helicity states, lepton helicity $\lambda$ is summed
up.}:
 \bea
  X_3 M_{\lambda_1\,\lambda_3}^{\Lambda_1 \Lambda_2}&=&
  A_2 \, V_{\lambda_1\lambda}^{\Lambda_1}(p_1,k_1) \,
         V_{\lambda\lambda_3}^{\Lambda_2}(p_1-k_1 +q, k_2) +
  \nn\\
  &+&
  q_\perp {B_2}_{\lambda_1\,\lambda_3}^{\Lambda_1 \Lambda_2} +
  \nn\\
  &+& 
  {\tilde A}_2 \frac{ \tilde V_{\lambda_1\lambda}^{\Lambda_1}(p_1, k_1)\,
  \tilde V_{\lambda\lambda_3}^{\Lambda_2} (p_1-k_1, k_2)}
  {4E_1^2(1-x_1)^2}\,
  \label{79}
\eea
with the scalars
\be
  A_2=  \frac{X_3}{a_1a_{12}} - \frac{1-x_1}{a_1b_2}+
  \frac{1}{b_{12} b_2}\,,\;\;\;
  {\tilde A}_2 =-\frac{X_3}{a_{12}} + \frac{1}{b_{12}}
  \label{80}
\ee
and the transverse 4--vector $B_2$
\bea
  &&{B_2}_{\lambda_1\,\lambda_3}^{\Lambda_1 \Lambda_2} =
  - X_3\,{2 {e}^{(\Lambda_2)\,*}_\perp\over a_1a_{12}}\,
  V_{\lambda_1 \lambda_3}^{\Lambda_1}(p_1,k_1)
  \nn\\
  &&\times
  \left( 1- {x_2\over 1-x_1}\, \delta_{\Lambda_2,- 2\lambda_3}
  \right) +
  \\
  &&+{2 {e}^{(\Lambda_1)\,*}_\perp\over  b_{12} b_2}\,
  V_{\lambda_1 \lambda_3}^{\Lambda_2}(p_1-k_1+q, k_2)\,\left( 1- x_1\,
  \delta_{\Lambda_1,- 2\lambda_1} \right)\,.
  \nn
  \label{81}
 \eea
It is not difficult to check that the quantities $A_2$ and
${\tilde A}_2$ vanish in the limit of small $q_\perp$:
 \be
  A_2\propto q_\perp\,, \;\;
  {\tilde A}_2 \propto q_\perp \,,
  \label{82}
 \ee
whereas $B_2$ is finite in this limit.

Let us stress that Eq.~(\ref{76}) together with relation
(\ref{79}) represents a very simple and compact expression for
all 16 helicity states, where all individual large (compared to
$ q_\perp$) contributions have been rearranged into finite
expressions.

\subsection{Explicit expressions for the helicity amplitudes}

Due to the parity conservation relation
\be
  {\cal{M}}_{-\lambda_1 \,\,-\lambda_3}^{-\Lambda_1 \;-\Lambda_2}
  = - (-1)^{\lambda_{1}+ \lambda_{3}} \left(
  {\cal{M}}_{\lambda_1 \,\lambda_3}^{\Lambda_1 \Lambda_2}
  \right)^* \,,
  \label{83}
\ee
there are only 8 independent helicity states of ${\cal M}$ among
the whole set of 16. We fix the choice of the independent
amplitudes by fixing the helicity of the initial electron to
$\lambda_1 = +1/2 =+$.
To find the amplitudes with given initial and final
helicities, we start from Eq.~(\ref{79}) and  substitute there
the expressions for vertices taken from  Eqs.~(\ref{44}),
(\ref{45}), (\ref{48})--(\ref{51}).

Using the complex combinations~(\ref{73}) we immediately obtain
the amplitudes $M$:
 \bea
  M_{++}^{++}&=&
  2 \left\{
  A_2 \frac{K_1^* R_2^*}{x_1 x_2 X_3}
  +\frac{K_1^* Q^*}{x_1 a_1 a_{12} }
  - \frac{Q^* R_2^*}{x_2 X_3 b_{12} b_2 } \right\} \,,
 \nn\\
  M_{++}^{--}&=& X_3 \left(M_{++}^{++} \right)^*\,,
  \label{84a} \\
  M_{++}^{-+}&=&-2 (1-x_1) 
   \nn\\
  &\times&\left(
  A_2 \frac{K_1 R_2^*}{x_1x_2X_3}
  +\frac{K_1 Q^*}{x_1 a_1 a_{12} }
  -\frac{Q R_2^*}{x_2 X_3  b_{12} b_2 }\right) \,,
  \label{84b} \\
  M_{++}^{+-}&=& {X_3\over (1-x_1)^2} \left(M_{++}^{-+}\right)^*+
  \nn\\
  &+&
  \frac{2}{1-x_1} \,\left( m^2 A_2 \frac{x_1 x_2}{X_3} -
  \tilde{A}_2\right)\,,
  \label{84c}\\
  M_{+-}^{+-}&=& 2 m x_1 \left( A_2 \frac{R_2}{x_2 X_3}
  +\frac{Q}{a_1 a_{12}}\right)\,,
  \nn\\
  M_{+-}^{-+}&=& 2 m \frac{x_2}{X_3}  \left( A_2 \frac{K_1}{x_1} -
  \frac{Q}{b_{12}   b_2}\right)\,,
  \nn\\
  M_{+-}^{--}&=&0\,,
  \label{84d}\\
  M_{+-}^{++}&=& -{1\over 1-x_1}\left(X_3\,M_{+-}^{+-}+
  M_{+-}^{-+}\right)^*\,.
  \label{84e}
\eea
All expressions are either identical [Eqs.(\ref{84a}),
(\ref{84b}), (\ref{84d})] to the amplitudes in
Ref.~\cite{KSSSh00} or can be transformed [Eqs.(\ref{84c}),
(\ref{84e})] after some algebra to those amplitudes.

As a result, we obtain all eight independent helicity  amplitudes
in the form\footnote{To clarify the notation we stress that in
Eqs.~(\ref{amplitudes}) with given polarizations the
operator ${\cal{P}}_{12}$ simply interchanges the indices
$1\leftrightarrow 2$ and
\begin{eqnarray*}
  {\cal{P}}_{12}A_2 = \frac{X_3}{a_2a_{12}} - \frac{1-x_2}{a_2b_1}
  + \frac{1}{b_1b_{12}} \,.
\end{eqnarray*}
}
\bea
  {\cal{M}}_{+\pm}^{++}&=& \left(1+{\cal{P}}_{12}\right)
  M_{+\pm}^{++}\,, \;\;
  {\cal{M}}_{++}^{--}= X_3  \left(\cal{M}_{++}^{++}\right)^* \,,
  \nonumber \\
  {\cal{M}}_{+\pm}^{+-}&=& M_{+\pm}^{+-} + {\cal{P}}_{12}
  M_{+\pm}^{-+}\,, \;\;
  {\cal{M}}_{+\pm}^{-+}={\cal{P}}_{12}
  {\cal{M}}_{+\pm}^{+-} \,,
  \\
  {\cal{M}}_{+-}^{--}&=&0 \,.
  \nn
  \label{amplitudes}
 \eea
The amplitudes are explicitly proportional to $Q$ or to the
functions $A_2$ and ${\tilde A}_2$. Therefore, they vanish
$\propto |{\mathbf{q}}_\perp|$ in the limit
$|{\mathbf{q}}_\perp| \to 0$.

The spin--flip amplitudes (with $\lambda_{1}=-\lambda_{3}$) are
proportional to the electron mass $ m$ and, therefore, they are
negligible compared to the spin non--flip ones for not too small
scattering angles
 \be
  \frac{m}{x_{1,2}E_1} \ll \theta_{1,2} \ll 1 \,, \; \;
  \frac{m}{X_3 E_1} \ll \theta_3 \ll 1 \,.
  \label{langle}
\ee
We also note the
explicit Bose symmetry  between the two photons in the
amplitudes~(\ref{amplitudes}):
\be
  {\cal{M}}^{\Lambda_1  \Lambda_2}_{\lambda_{1} \lambda_{3}}
  (K_1,x_1; K_2,x_2)=
  {\cal{M}}^{\Lambda_2  \Lambda_1}_{\lambda_{1} \lambda_{3}}
  (K_2,x_2; K_1,x_1) \,.
\ee

\section{Impact factor for the multiple bremsstrahlung 
$e(p_1)+ \gamma^*(q) \to e(p_3)+ \gamma(k_1)+\ldots+ \gamma(k_n)$}

\begin{figure}[!htb]
  \begin{center}
  \unitlength=2.00mm
  \begin{picture}(45.00,13.00)(-1.00,2.50)
    \put(20.80, 8.50){\line(1,0){9.0}}
    \put(20.80,8.50){\vector(1,0){1.50}}
    \put(23.80,8.50){\vector(1,0){1.5}}
    \put(27.80,8.50){\vector(1,0){1.5}}
    \put(26.80,8.50){\vector(1,0){10.0}}
    \put(31.80,8.50){\vector(1,0){1.5}}
    \put(1.50,7.00){\makebox(0,0)[cc]{$p_1$}}
    \put(16.50,7.00){\makebox(0,0)[cc]{$p_3$}}
    \put(30.80, 2.50){\line(0,1){2.00}}
    \put(30.80,5.00){\line(0,1){2.00}}
    \put(30.80,7.50){\line(0,1){1.00}}
    \put(30.80,5.00){\vector(0,1){1.40}}
    \put(29.30,5.50){\makebox(0,0)[cc]{$q$}}
    \put(39.80,13.50){\line(1,1){1.8}}
    \put(37.30,11.00){\line(1,1){1.8}}
    \put(34.80,8.50){\line(1,1){1.8}}
    \put(37.30,11.0){\vector(1,1){1.8}}
    \put(35.90,11.20){\makebox(0,0)[cc]{${k_3}$}}
    \put(31.80,13.50){\line(1,1){1.8}}
    \put(29.30,11.00){\line(1,1){1.8}}
    \put(26.80,8.50){\line(1,1){1.8}}
    \put(29.30,11.0){\vector(1,1){1.8}}
    \put(27.90,11.20){\makebox(0,0)[cc]{${k_2}$}}
    \put(27.80,13.50){\line(1,1){1.8}}
    \put(25.30,11.00){\line(1,1){1.8}}
    \put(22.80,8.50){\line(1,1){1.8}}
    \put(25.30,11.0){\vector(1,1){1.8}}
    \put(23.90,11.20){\makebox(0,0)[cc]{${k_1}$}}
    \put(1.00, 8.50){\line(1,0){9.0}}
    \put(1.00,8.50){\vector(1,0){1.50}}
    \put(4.00,8.50){\vector(1,0){1.5}}
    \put(8.00,8.50){\vector(1,0){1.5}}
    \put(7.00,8.50){\vector(1,0){10.0}}
    \put(12.00,8.50){\vector(1,0){1.5}}
    \put(15.00, 2.50){\line(0,1){2.00}}
    \put(15.00,5.00){\line(0,1){2.00}}
    \put(15.00,7.50){\line(0,1){1.00}}
    \put(15.00,5.00){\vector(0,1){1.40}}
    \put(13.50,5.50){\makebox(0,0)[cc]{$q$}}
    \put(16.00,13.50){\line(1,1){1.8}}
    \put(13.50,11.00){\line(1,1){1.8}}
    \put(11.00,8.50){\line(1,1){1.8}}
    \put(13.5,11.0){\vector(1,1){1.8}}
    \put(14.10,11.20){\makebox(0,0)[cc]{${k_3}$}}
    \put(12.00,13.50){\line(1,1){1.8}}
    \put(9.50,11.00){\line(1,1){1.8}}
    \put(7.00,8.50){\line(1,1){1.8}}
    \put(9.5,11.0){\vector(1,1){1.8}}
    \put(8.10,11.20){\makebox(0,0)[cc]{${k_2}$}}
    \put(8.00,13.50){\line(1,1){1.8}}
    \put(5.50,11.00){\line(1,1){1.8}}
    \put(3.00,8.50){\line(1,1){1.8}}
    \put(5.50,11.0){\vector(1,1){1.8}}
    \put(4.10,11.20){\makebox(0,0)[cc]{${k_1}$}}
  \end{picture}
  \begin{picture}(74.00,14.00)(39.50,2.50)
    \put(40.60, 8.50){\line(1,0){9.0}}
    \put(40.60,8.50){\vector(1,0){1.50}}
    \put(43.60,8.50){\vector(1,0){1.5}}
    \put(47.60,8.50){\vector(1,0){1.5}}
    \put(46.60,8.50){\vector(1,0){10.0}}
    \put(51.60,8.50){\vector(1,0){1.5}}
    \put(46.60, 2.50){\line(0,1){2.00}}
    \put(46.60,5.00){\line(0,1){2.00}}
    \put(46.60,7.50){\line(0,1){1.00}}
    \put(46.60,5.00){\vector(0,1){1.40}}
    \put(45.00,5.50){\makebox(0,0)[cc]{$q$}}
    \put(55.60,13.50){\line(1,1){1.8}}
    \put(53.10,11.00){\line(1,1){1.8}}
    \put(50.60,8.50){\line(1,1){1.8}}
    \put(53.10,11.00){\vector(1,1){1.8}}
    \put(50.70,11.20){\makebox(0,0)[cc]{${k_2}$}}
    \put(59.60,13.50){\line(1,1){1.8}}
    \put(57.10,11.00){\line(1,1){1.8}}
    \put(54.60,8.50){\line(1,1){1.8}}
    \put(57.10,11.00){\vector(1,1){1.8}}
    \put(55.70,11.20){\makebox(0,0)[cc]{${k_3}$}}
    \put(47.60,13.50){\line(1,1){1.8}}
    \put(45.10,11.00){\line(1,1){1.8}}
    \put(42.60,8.50){\line(1,1){1.8}}
    \put(45.10,11.00){\vector(1,1){1.8}}
    \put(43.70,11.20){\makebox(0,0)[cc]{${k_1}$}}
    \put(61.40,8.50){\line(1,0){9.0}}
    \put(60.40,8.50){\vector(1,0){1.50}}
    \put(63.40,8.50){\vector(1,0){1.5}}
    \put(67.40,8.50){\vector(1,0){1.5}}
    \put(66.40,8.50){\vector(1,0){10.0}}
    \put(71.40,8.50){\vector(1,0){1.5}}
    \put(63.40, 2.50){\line(0,1){2.00}}
    \put(63.40,5.00){\line(0,1){2.00}}
    \put(63.40,7.50){\line(0,1){1.00}}
    \put(63.40,5.00){\vector(0,1){1.40}}
    \put(62.00,5.50){\makebox(0,0)[cc]{$q$}}
    \put(75.00,13.50){\line(1,1){1.8}}
    \put(72.50,11.00){\line(1,1){1.8}}
    \put(70.00,8.50){\line(1,1){1.8}}
    \put(72.50,11.00){\vector(1,1){1.8}}
    \put(71.10,11.20){\makebox(0,0)[cc]{${k_2}$}}
    \put(71.00,13.50){\line(1,1){1.8}}
    \put(68.50,11.00){\line(1,1){1.8}}
    \put(66.00,8.50){\line(1,1){1.8}}
    \put(68.50,11.00){\vector(1,1){1.8}}
    \put(67.10,11.20){\makebox(0,0)[cc]{${k_1}$}}
    \put(79.00,13.50){\line(1,1){1.8}}
    \put(76.50,11.00){\line(1,1){1.8}}
    \put(74.00,8.50){\line(1,1){1.8}}
    \put(76.50,11.0){\vector(1,1){1.8}}
    \put(75.10,11.20){\makebox(0,0)[cc]{${k_3}$}}
  \end{picture}
  \end{center}
  \caption{Feynman diagrams for the impact factor related to the
  triple bremsstrahlung, diagrams with the
  exchange of the final photons have to be added.}
  \label{fig:14}
\end{figure}
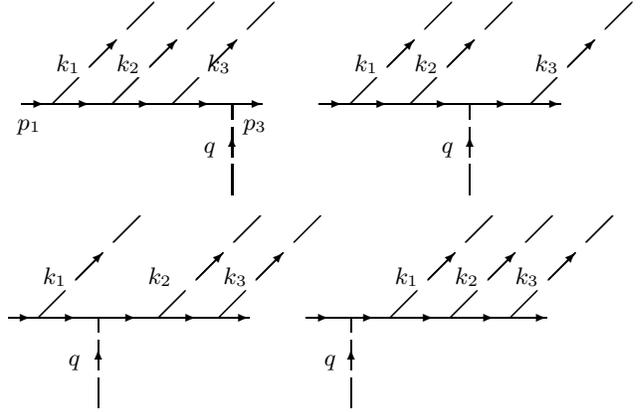

The generalization of the results obtained in Sects. 4 and 5 to
the bremsstrahlung of $n$ photons can be done straightforwardly.
To demonstrate this, we consider the case $n=3$
(Fig.~\ref{fig:14})
which clearly shows all nontrivial points of the multiple
bremsstrahlung.

The helicity states of the initial and
final electrons $\lambda_{1,3}$ and of the final photons
$\Lambda_{1,2,3}$ are indicated explicitly in the impact factor:
 \bea
  J_{1}&=& \sqrt{2} \, (4\pi\alpha)^{2}\,
  \\
  &\times&X_3\,
  {\cal M}_{\;\;\lambda_1\, \lambda_3}^{\Lambda_1\, \Lambda_2\,\Lambda_3}\,
  (x_1, x_2, x_3, k_{1\perp},  k_{2\perp}, k_{3\perp},
  p_{3\perp}) \; \Phi\,. \nn
  \label{92}
 \eea
where $\Phi$ has the form~(\ref{71a}). The quantities $J_{1}$
and ${\cal M}$ do not depend on $s$, but depend only on the
energy fractions
 $$
  x_{1,2,3} =\omega_{1,2,3}/E_1, \;\; X_3 = E_3/E_1\,, \;\;
  x_1+x_2+x_3+X_3=1
$$ and on the transverse momenta of the final particles in the
first jet with
 $$
q_\perp= \sum_{i=1}^3 k_{i\perp}+p_{3\perp} \,.
 $$
For the denominators of the propagators in the diagrams of
Fig.~\ref{fig:14} we use the notations ($i,j=1,2,3$)
 \bea
  \label{93}
  a_{i}&=&-(p_1-k_i)^2+m^2 \,, \nn\\
  b_{i}&=&(p_3+k_i)^2-m^2\,,\nn\\
  a_{ij}&=&-(p_1-k_i-k_j)^2+m^2 \,,\\
  b_{ij}&=&(p_3+k_i+k_j)^2-m^2 \,, \nn\\
  a_{123}&=&-(p_1-k_1-k_2-k_3)^2+m^2 \,,\nn\\
  b_{123}&=&(p_3+k_1+k_2+k_3)^2-m^2\,.\nn
 \eea

Just as in Sect. 5.2, we obtain
\bea
  {\cal M}&=& (1+ {\cal P}) ( M +\tilde M)\,,
  \\
  {\cal P}&=& {\cal P}_{12} +{\cal P}_{23}+
  {\cal P}_{13}+{\cal P}_{13} {\cal
  P}_{12}+{\cal P}_{13} {\cal P}_{23}\,,
  \nn
  \label{94}
\eea
 \bea
  X_3 M &=& {X_3\over a_1 a_{12}a_{123}}\, V(p_1, k_1)\nn\\
  &\times&
  V(p_1-k_1, k_2)\,V(p_1-k_1-k_2, k_3)- 
  \nn\\
  &-& {1-x_1-x_2\over a_1a_{12} b_3}\, V(p_1, k_1) \nn\\
  &\times&
  V(p_1-k_1, k_2)V(p_1-k_1-k_2+q, k_3)+
  \nn\\
  &+&{1-x_1\over a_1 b_{23} b_3}\, V(p_1, k_1) \\
  &\times& V(p_1-k_1+q, k_2)
  V(p_1-k_1-k_2+q, k_3)-
  \nn\\
  &-&{1\over b_{123} b_{23} b_3}\, V(p_1+q, k_1)\nn\\
  &\times& V(p_1-k_1+q, k_2)
  V(p_1-k_1-k_2+q, k_3)\,,\nn
  \label{95}
\eea
 \bea
  X_3 \tilde M &=&- {X_3\over a_{12}a_{123}}\,
  { \tilde V(p_1, k_1)\, \tilde
  V(p_1-k_1, k_2)\over 4(E_1-\omega_1)^2} 
  \nn\\
  &\times& V(p_1-k_1-k_2, k_3)   -
  \nn\\
  &-& {X_3\over a_{12}a_{123}}\,  V(p_1, k_1) 
  \nn\\
  &\times&  
  { \tilde V(p_1-k_1,
  k_2) \tilde V(p_1-k_1-k_2, k_3)\over
  4(E_1-\omega_1-\omega_2)^2}+
  \nn\\
  &+& {1-x_1-x_2\over a_{12}b_{3}}\, { \tilde V(p_1, k_1)\, \tilde
  V(p_1-k_1, k_2)\over 4(E_1-\omega_1)^2} 
  \nn\\
  &\times& V(p_1-k_1-k_2+q, k_3)+
  \nn\\
  &+& {1-x_1\over a_{1}b_{23}}\,  V(p_1, k_1)
  \\
  &\times&{\tilde
  V(p_1-k_1+q, k_2) \tilde V(p_1-k_1-k_2+q, k_3)\over
  4(E_1-\omega_1-\omega_2)^2}- 
  \nn\\
  &-& {1\over b_{123}b_{3}}\, { \tilde V(p_1+q, k_1)\, \tilde
  V(p_1-k_1+q, k_2)\over 4(E_1-\omega_1)^2} 
  \nn\\
  &\times& V(p_1-k_1-k_2+q, k_3)-
  \nn\\
  &-& {1\over b_{123}b_{23}}\,  V(p_1+q, k_1)
  \nn\\
  &\times& { \tilde
  V(p_1-k_1+q, k_2) \tilde V(p_1-k_1-k_2+q, k_3)\over
  4(E_1-\omega_1-\omega_2)^2}\,.
  \nn
  \label{96}
\eea

To show that $J_1$ vanishes as $\propto q_\perp$ in the limit
$q_\perp \to 0$, we follow the same line of action as in the
previous section. This gives
\bea
  X_3 M&=&  A_3\,V(p_1,k_1)\,  V(p_1-k_1,k_2)V(p_1-k_1-k_2,k_3)+
  \nn\\
  &+&
  q_\perp B_3\,,
  \nn\\
  X_3 \tilde M &=&{\tilde A}_3\, { \tilde V(p_1, k_1)\,
  \tilde V(p_1-k_1, k_2)\over
  4(E_1-\omega_1)^2} V(p_1-k_1-k_2, k_3)+
  \nn\\
  &+& {\tilde A}'_3\,  V(p_1, k_1)\,{ \tilde V(p_1-k_1, k_2)
  \tilde V(p_1-k_1-k_2, k_3)\over 4(E_1-\omega_1-\omega_2)^2} +
  \nn\\
   &+&
  q_\perp {\tilde B}_3
  \label{97}
\eea
where
\bea
  A_3&=& {X_3\over a_1 a_{12}a_{123}}\, - {1-x_1-x_2\over
  a_1a_{12} b_3}\,+{1-x_1\over a_1 b_{23} b_3}\, -{1\over b_{123}
  b_{23} b_3}\,,\nn\\
  {\tilde A}_3&=&
  - {X_3\over a_{12}a_{123}}+{1-x_1-x_2\over a_{12}b_{3}}-{1\over
  b_{123}b_{23}}\,,\nn\\
  {\tilde A}'_3&=&
  - {X_3\over a_{12}a_{123}}+{1-x_1\over a_{1}b_{23}}-{1\over
  b_{123}b_{23}}
  \label{98}
 \eea
and the 4--vectors $B_3$ and ${\tilde B}_3$ can be easily found
from Eqs.~(\ref{95})--(\ref{98}). The quantities $A_3,\, {\tilde
A}_3$ and ${\tilde A}'_3$ vanish in the limit of small $q_\perp$
 \be
  A_3          \propto q_\perp \,, \;\;
  {\tilde A}_3 \propto q_\perp \,,\;\;
  {\tilde A}'_3\propto q_\perp \,,
  \label{99}
 \ee
the transverse 4--vectors $B_3$ and ${\tilde B}_3$ remain finite
in this limit.

Again, Eq.~(\ref{94}) with relations  (\ref{97}) is a very
simple and compact expression for all 32 helicity states where
all individual large (compared to $ q_\perp$) contributions have
been cancelled.

\section{Some general properties of
bremsstrahlung impact factors}

We discuss now some general properties of impact factors for the
emission of real photons using mainly the double bremsstrahlung
as example.
For that case Eqs.~(\ref{71}), (\ref{76}) and
(\ref{79}) define  a simple, compact and transparent expression
for the vertex factor which allows to obtain immediately all
general properties obtained in Ref.~\cite{KSSSh00} only after
lengthy calculations. All those properties are directly related
to the corresponding properties of vertices discussed in Sect.
3.3.

 1)
Bremsstrahlung amplitudes or impact factors with a maximal
change of helicities are absent since in this case at least one
transition vertex has to appear with a maximal change of its
helicities ($=2$) as well. Thus we have
\be
  {\cal M}=0\; \mbox{ for } \; \max |\Delta \lambda| = n+1
  \label{36c}
\ee
where $\Delta \lambda= \sum_{i=1}^n \Lambda_i + \lambda_3
-\lambda_1$ is the  change of helicity in the transition from
the first initial lepton to the first jet. In the case of the
double bremsstrahlung, this corresponds to
$$
  {\cal M}_{+-}^{--} = {\cal M}_{-+}^{++}=0\,.
$$

 2)
If one of the final particles in the jet (including the final
lepton) becomes hard ($\omega_i\to E_1$ or $E_3 \to E_1$) the
sign of the helicity of the initial lepton coincides with that
of the helicity of the hard final particle. This is the
consequence of properties 2) and 3) discussed in Sect. 3.3.

 3)
In HNC amplitudes the sign of the helicity of at least one final
photon has to  coincide with the sign of the initial lepton
helicity
\be
  {\cal M}_{\lambda_1\; -\lambda_1}^{\Lambda_1\cdots \Lambda_n} \propto
  \delta_{\Lambda_i, 2 \lambda_1} \,.
\ee

 4)
The dependence of the whole amplitude $M$ on complex parameters
of the form $z$ and $z^*$ defined in (\ref{48}) can be easily
reproduced for  the whole amplitude using Eq. (\ref{56}).

 5)
As can be seen from Eqs. (\ref{47}), (\ref{50}) and (\ref{51}),
the $\tilde V$ vertex may contribute only if the electron line
connects  two vertices  with the emission of real photons. Both
these adjacent vertices are of HNC type transitions so that the
original lepton helicity is reestablished after passing these
two vertices going along the lepton line. In our example, this
happens for the two independent amplitudes (contributions
including the $A_2'$ factor):
$$
  {\cal M}_{++}^{+-} = \left( {\cal M}_{--}^{-+} \right)^*\,,
  \;\;\; {\cal M}_{++}^{-+} = \left( {\cal M}_{--}^{+-}
  \right)^*\,.
  \label{29c}
$$
Since the number of real photons is $2$ in that case, initial
and final lepton helicities have to  coincide for those
amplitudes.

 6)
It is known that for soft photons (approximation of classical
currents) the bremsstrahlung matrix element factorizes into a
term responsible for the soft photon times an amplitude without
the soft photon. In our approach this can can be easily realized
using the following arguments. The form of a vertex in the soft
photon limit is given by  Eq.~(\ref{54}). Furthermore, a virtual
electron propagator close to a soft photon might have an
infrared singularity only, if the soft photon is either at the
beginning or the end of the electron line in a Feynman diagram.
This has also the consequence that only those diagrams can
contribute to the soft photon limit. Therefore, for a soft
photon at the beginning of the electron line we have the vertex
$$
  V(p_1,k_1)\to -\frac{2}{x_1}\,
  (e_\perp^{(\Lambda_1)\:*} k_{1\perp})\;
  \delta_{\lambda_1\lambda}=2
  (e^{(\Lambda_1)\:*}p_1)\;\delta_{\lambda_1\lambda}
$$
and the denominator of the corresponding electron propagator
$$
  (p_1-k_1)^2-m^2 =-2p_1k_1
$$
(and analogously for the soft photon at the end of the electron
line). The remaining part of the amplitude is then taken at
$k_1=0$ and represents the impact factor for $n-1$
bremsstrahlung photons. As a result, we get the factorization
property for the impact factors (assuming the first photon being
soft):
\bea
  &J_1&\left( e_{\lambda_1}(p_1)+ \gamma^*(q) \to
  e_{\lambda_3}(p_3) + \sum_{i=1}^{n}
  \gamma_{\Lambda_i}(k_i) \right) \to \nn \\  
  &\to&\sqrt{4 \pi
  \alpha} 
  \left(\frac{e^{(\Lambda_1)\:*} p_1}{p_1 k_1}
     -\frac{e^{(\Lambda_1)\:*} p_3}{p_3 k_1}   \right) 
   \\
  &\times&
  J_1 \left( e_{\lambda_1}(p_1)+ \gamma^*(q) \to e_{\lambda_3}(p_3)
  + \sum_{i=2}^{n} \gamma_{\Lambda_i}(k_i)\right)\,.
  \nn
\eea
The generalization to $m$ soft (first) photons out of $n$
bremsstrahlung photons is obvious:
\bea
  &J_1&\left(e_{\lambda_1}(p_1)+ \gamma^*(q) \to 
  e_{\lambda_3}(p_3) + \sum_{i=1}^{n}
  \gamma_{\Lambda_i}(k_i)\right) \to 
  \nn \\
  &\to&\left(4 \pi
  \alpha\right)^{\frac{m}{2}} 
   \,
  \left\{\prod_{i=1}^m
      \left(\frac{e^{(\Lambda_i)\:*} p_1}{p_1 k_i}
     -\frac{e^{(\Lambda_i)\:*} p_3}{p_3 k_i}   \right)\right\}
   \\
  &\times& J_1\left(e_{\lambda_1}(p_1)+ \gamma^*(q) \to e_{\lambda_3}(p_3)
  + \sum_{i=m}^{n} \gamma_{\Lambda_i}(k_i)\right)\,.
  \nn
\eea

 7)
For a process with the emission of $n$ real photons we have the
relation between the impact factors for initial positron and
electron:
\bea
  &&J_1 (e^{+}_{\lambda_1}+\gamma^* \to e^{+}_{\lambda_3}+
  \gamma_{\Lambda_1}+ ...+\gamma_{\Lambda_n })=
  \\
  &&=
  (-1)^{n+1} \,
  J_1 (e^{-}_{\lambda_1}+\gamma^* \to e^{-}_{\lambda_3}+
  \gamma_{\Lambda_1}+ ...+\gamma_{\Lambda_n })\,.
  \nn
  \label{emtoep} 
\eea
This may be easily proven repeating the arguments given for the
single bremsstrahlung in Sect. 4.

 8)
Let us consider the connection between the impact factor $J_1$
for the first jet discussed in Sect. 3---6 and the impact factor
$J_2$ for the second jet. If the impact factor $J_1$ is related
to the process
$$
  e_{\lambda_1} (p_1)+\gamma^*(q) \to e_{\lambda_3}(p_3)+
 \gamma_{\Lambda_1}(k_1) + \ldots +\gamma_{\Lambda_n }(k_n)\,,
$$
it is a function of the following parameters
$$
  J_1 \equiv J_1 (\lambda_1;\, \lambda_3, X_3, p_{3\perp} ;\,
  \Lambda_1, x_1, k_{1\perp} ;\, \ldots ;\,
  \Lambda_n, x_n, k_{n\perp} )
$$
where $ X_3=E_3/E_1$ and $x_j = \omega_j/E_1$. The impact
factor $J_2$, related to the process
$$
  e_{\lambda_2} (p_2)+\gamma^*(-q) \to e_{\lambda_4}(p_4)+
  \gamma_{\tilde{\Lambda}_1}(\tilde k_1) +
  \ldots+\gamma_{\tilde{\Lambda}_n }(\tilde k_n) \,,
$$
depends on the parameters
$$
  J_2 \equiv J_2 (\lambda_2;\, \lambda_4, Y_4, p_{4\perp} ;\,
  \tilde{\Lambda}_1, \tilde y_1, \tilde k_{1\perp} ;\, \ldots ;\,
  \tilde{\Lambda}_n, \tilde y_n, \tilde k_{n\perp} )
$$
where $Y_4=E_4/E_2$ and $\tilde y_j = \tilde \omega_j/E_2$.
Any 4--vector $\tilde k = (\omega,\, {\bf k}_\perp ,\,
-k_z)$ for a particle in the second jet can be obtained from the 
4--vector $k = (\omega,\, {\bf k}_\perp ,\, k_z)$ for a
particle in the first jet by spatial inversion and further
rotation by an angle $\pi$ around the new $z$-axis. Since this 
operation changes the signs of helicities of leptons and photons, 
the impact factor $J_2$ is derived from $J_1$ by the following 
substitution rule:  
\bea
  J_2 &=&J_1 (-\lambda_2;\, -\lambda_4, Y_4, p_{4\perp} ;\,
  - \tilde{\Lambda}_1, \tilde y_1, \tilde k_{1\perp} ;\, \ldots
  \nn\\
  &&\dots
  ;\, -\tilde{\Lambda}_n, \tilde y_n, \tilde k_{n\perp} )\,.
  \label{J12}
\eea

\section{Summary}

In the present paper we have formulated a new
effective method to calculate all helicity amplitudes for
bremsstrahlung jet--like QED processes at tree level.

The jet kinematic conditions here considered (\ref{2}) provide 
the main contribution to the total cross sections of these processes at 
high energy. Within this kinematics, it is possible to obtain simple 
expressions of helicity amplitudes with an accuracy defined by (\ref{3}). 
In this region, these amplitudes can be presented in the simple factorized 
form (\ref{4}), where the impact 
factors $J_1$ or $J_2$ are proportional to the scattering amplitudes of 
the first or second initial lepton with the virtual exchanged photon.

The main advantage of our method consists in using simple universal 
``building blocks'' --- transition vertices with {\it real} leptons ---
which are matrices with respect to lepton helicities.
Those vertices replace efficiently the spinor structure involving 
leptons of small virtuality in the impact factors, making the calculations 
short and transparent for any final helicity state. In the calculations 
we exploit a convenient decomposition of all 4--momenta
of the reaction into large and small components involving Sudakov (or 
light-cone) variables.

The vertices themselves or their allowed combinations with 
well--defined prefactors (see discussion in Sect.~3.1) 
are finite in the high energy
limit  $s\to \infty$.
In the case of bremsstrahlung we have found that  
only three nonzero transition vertices are required.
The calculation of the vertices can be conveniently performed using
the spinor or chiral representation of bispinors and $\gamma$--matrices. The
properties of the vertices, discussed in Sect.~3.3, determine all 
nontrivial general properties of the helicity 
amplitudes described in Sect.~7.

By construction, the impact factors are finite in the high energy limit and 
depend only on energy fractions and transverse momenta of particles in the 
final jet, and on helicities of all real photons and leptons.

In Sections 4--6 we have calculated the impact factors
for single, double and triple bremsstrahlung, following the
same principles. In a first step, we use the allowed	
vertices to write down the corresponding impact factors, see Eqs.
(\ref{59}), (\ref{76}), (\ref{77}) and (\ref{94})--(\ref{96}).
In the next step, we use gauge invariance with
respect to the virtual photon of 4--momentum $q$ and rearrange
impact factors into a form in which all individual large (compared to $
q_\perp$) contributions have been cancelled, see Eqs.
(\ref{64}), (\ref{67}), (\ref{79}), (\ref{amplitudes}) and
(\ref{97}).
Let us stress here again that the known results for
bremsstrahlung helicity amplitudes to order $e^2$ and $e^3$  are now 
obtained almost immediately using this new method, while handling the 
spinor structure directly leads to cumbersome and tedious calculations in
the case, for instance, of double photon bremsstrahlung. 
The result of order $e^4$ for the triple bremsstrahlung in one direction 
is completely new.

We have also defined rules to go over 
from impact factors with initial electrons to
those with positrons [see Eq.~(\ref{emtoep})] and 
from the impact factors for the first
jet to that for second jet [Eq.~(\ref{J12})].

Those rules together with the found impact factors allows us to give a
complete analytic and compact description of all helicity amplitudes in 
$e^-e^\pm$ scattering with the emission of up to three photons in one 
lepton direction, where in the last case 
$2^5 \times 2^5$ different helicity amplitudes  are involved.

Since by construction individual large contributions (compared to 
$q_\perp$) have been rearranged into finite expressions,
the expressions obtained for the amplitudes are 
very convenient for numerical calculations of various cross sections.

Until now we have formulated our new method only for the case of photon 
bremsstrahlung from
leptons. A next paper will be devoted to QED
processes with production of lepton pairs~\cite{CSS2}.

\section*{Acknowledgements}

We are grateful to S.~Brodsky, V.~Fadin, I.~Ginzburg,
L.~McLerran and A.~Vainshtein for useful discussions. This work
is supported in part by INTAS (code 00-00679), RFBR (code
00-02-17592) and by S.--Petersburg grant (code E00-3.3-146).

\section*{Appendix}

In the Appendix we collect some useful formulae about {\it
spinor or chiral} representation (see, for example,
Ref.~\cite{KS} and text-book~\cite{BLP} \S 20, 21, 26).

We start with the {\it standard} representation in which an
electron with momentum $\bf p$, energy $E=\sqrt{{\bf p}^2+m^2}$
and helicity $\lambda=\pm1/2$ is described by the bispinor
\begin{eqnarray*}
  u^{(\lambda)}_{\mathbf p}&=& \left(
  \begin{array}{c}
    \sqrt{E+m}\,w^{(\lambda)} ({\mathbf n})\\
     2\lambda\,\sqrt{E-m}\, w^{(\lambda)} ({\bf n})
  \end{array} \right)\,, 
  \\
  {\mathbf n}&=&{{\mathbf p}\over \mid{\mathbf
  p}\mid}=(\sin\theta\cos\varphi,\,   
  \sin\theta\sin\varphi,\,\cos\theta)\,.
\end{eqnarray*}
The two-component spinors $w^{(\lambda )} ({\bf n})$ obey
the equations
$$
  (\mbox{\boldmath $ \sigma$}{\bf n })\,w^{(\lambda
  )} ({\bf n})=2\lambda\, w^{(\lambda )} ({\bf n})\,,\;\;
  w^{(\lambda )+} ({\bf n})\,w^{(\lambda ')}({\bf n})=\delta
  _{\lambda\lambda'}
$$
and have the form
\begin{eqnarray*}
  w^{(1/2)} ({\bf n})&=&\left(
  \begin{array}{c}
    {\mathrm e}^{-{\rm i}\varphi /2}\cos{\theta \over 2}  \\
    {\mathrm e}^{{\rm i} \varphi /2}\sin{\theta \over 2}
  \end{array} \right) \,, \\
  w^{(-1/2)}({\bf n})&=&\left(
  \begin{array}{c}
    -{\mathrm e}^{-{\rm i}\varphi /2}\sin{\theta \over 2} \\
     {\mathrm e}^{{\rm i}\varphi /2}\cos{\theta\over 2}
  \end{array} \right)
\end{eqnarray*} 
with properties
$$
 \sigma_y w^{(\lambda)\,*} ({\bf n}) = 2\lambda {\rm i}
  w^{(-\lambda)} ({\bf n})\,,\;\;
  w^{(\lambda)} (-{\bf n})= {\rm i} w^{(-\lambda)} ({\bf n})
$$
(here \mbox{\boldmath $\sigma$} are the Pauli matrices). The
normalization conditions are
$$
  \bar u^{(\lambda )}_{\bf p}
  u^{(\lambda ')}_{\bf p}=2m\,\delta _{\lambda\lambda '}\,,\qquad
  \sum _\lambda u^{(\lambda )}_{\bf p} \bar u ^{(\lambda )}_{\bf
  p}=\hat p +m \,.
$$
For the initial electron with momentum ${\bf p}_1$ (${\bf p}_2$)
along (opposite) the $z$-axis we use $\theta =0$ ($\theta
=\pi$). For the final electron with momentum ${\bf p}_3$ in the
first jet we use $\theta =\theta_3$ and for the final electron with
${\bf p}_4$ in the second jet $\theta =\pi-\theta_4$.

The Dirac matrices in the standard representation are defined as
$$
  \gamma^0=\left(
  \begin{array}{cc}
    1 & 0 \\
    0& -1
  \end{array}
  \right) \,,\; \;
  \mbox{\boldmath $\gamma$}=\left(
  \begin{array}{cc}
    0& \mbox{\boldmath $\sigma$} \\
    -\mbox{\boldmath $\sigma$}& 0
  \end{array}
  \right) \,,\;\;
  \gamma^5=\left(
  \begin{array}{cc}
    0 & -1 \\
    -1&  0
  \end{array}
  \right) \,.
$$

A positron with momentum $\bf p$, energy $E=\sqrt{{\bf
p}^2+m^2}$ and helicity $\lambda$ is described by the bispinor
$$
  v^{(\lambda)}_{\mathbf p}= C
  \left(\bar{u}^{(\lambda)}_{\mathbf p}\right)^T\,,
$$
with the charge conjugation matrix
$$
   C=\gamma^2 \gamma^0\,, \;\;
   C= -C^T= C^{-1}\,,\;\; C^{-1} \gamma_\mu C = - \gamma_\mu^T\,.
$$
Therefore, for a positron we get the bispinor
$$
  v^{(\lambda)}_{\bf p}={\rm i} { \sqrt{E-m}\, w^{(-\lambda)}
  ({\bf n}) \choose -2\lambda\,\sqrt{E+m}\,w^{(-\lambda)} ({\bf
  n})} \,,
$$
with the normalization conditions
$$
 \bar v^{(\lambda)}_{\bf p}v^{(\lambda')}_{\bf p}=-2
  m\,\delta _{\lambda\lambda'} \,, \;\;\; \sum _\lambda
  v^{(\lambda)}_{\bf p}\bar v^{(\lambda)}_{\bf p}=\hat p-m \,.
 $$
At high energies the bispinors $u$ and $v$ in the standard
representation have top and bottom components of the same order,
$\sim \sqrt{E}$, with relative corrections $\sim m/E$.

A more simple and convenient structure of bispinors can be found
in the spinor or chiral representation, the transition to which
is given by the matrix
 \begin{eqnarray*}
  U=U^{-1}= {1\over \sqrt{2}} \left(\gamma^0-\gamma^5\right)=
  {1\over \sqrt{2}}\, \left(
  \begin{array}{cc}
  1 & 1 \\
  1& -1
  \end{array}
  \right)\,.
  \label{ap1}
\end{eqnarray*}
In the spinor representation the electron bispinor is
\begin{eqnarray*}
  Uu^{(\lambda)}_{\bf p}= {1\over \sqrt{2}}\, \left(
  \begin{array}{c}
  \left(\sqrt{E+m}+2\lambda  \sqrt{E-m}\;\right)  \,
  w^{(\lambda)} ({\bf n})
  \\
  [3mm] \left(\sqrt{E+m}-2\lambda  \sqrt{E-m}\;\right)  \,
  w^{(\lambda)} ({\bf n})
  \end{array}
  \right)\,.
  \label{ap2}
\end{eqnarray*}
At high energies $E\gg m$ this bispinor has a large top
component $\sim\sqrt{E}$ and a small  bottom component
$\sim(m/\sqrt{E})$ for $\lambda =+ 1/2$ and vice versa for
$\lambda =- 1/2$ what  is very useful for analysis. Furthermore,
in that representation the corrections to the high energy
asymptotics
\begin{eqnarray*}
  Uu^{(\lambda)}_{\bf p} \approx {\sqrt{m}}\, \left(
  \begin{array}{c}
   (2E/m)^{\lambda}  \, w^{(\lambda)} ({\bf n})
   \\
  [3mm]
   (2E/m)^{-\lambda}    \,
   w^{(\lambda)} ({\bf n})
  \end{array}
  \right)
  \label{ap3}
\end{eqnarray*}
is of the relative order of $m^2/E^2$. The approximate formulae
for the positron bispinor in that representation are

\begin{eqnarray*}
  Uv^{(\lambda)}_{\bf p} &\approx& 2{\rm i}
  \lambda\,{\sqrt{m}}\, \left(
  \begin{array}{c}
     -(2E/m)^{-\lambda}  \, w^{(-\lambda)} ({\bf n})
    \\
    [3mm]
    (2E/m)^{\lambda}\, w^{(-\lambda)} ({\bf n})
  \end{array}
  \right),\\
   Uv^{(\lambda)}_{-{\bf p}} &\approx& 2
  \lambda\,{\sqrt{m}}\, \left(
  \begin{array}{c}
   (2E/m)^{-\lambda}  \, w^{(\lambda)} ({\bf n})
   \\
  [3mm]
   -(2E/m)^{\lambda}\,
   w^{(\lambda)} ({\bf n})
  \end{array}
  \right).
\end{eqnarray*}

Omitting terms of the order of $\theta^2$, we obtain the
following simple expression for the two--component spinor:
\begin{eqnarray*}
   w^{(\lambda=+1/2)} ({\bf n})&=&\left(
  \begin{array}{c}
  1  \\ a
  \end{array} \right)\, {\mathrm e}^{-{\rm i}\lambda \varphi}
   \,, \\
  w^{(\lambda=-1/2)} ({\bf n})&=&\left(
  \begin{array}{c}
  a \\ 1
  \end{array} \right)\, {\mathrm e}^{-{\rm i}\lambda \varphi}
  \label{aa5}
\end{eqnarray*}
where
$$
  a = \lambda \, \theta\, {\rm e}^{2{\rm i} \lambda \varphi}=
  -{1\over \sqrt{2}\,E} \; p_\perp \, e^{(-2\lambda)*}_\perp
  \label{aa6}
$$
and the 4--vector $e_\perp^{(\Lambda)}$ is given in (\ref{16}).

To calculate the vertices (\ref{36}), (\ref{37}) and (\ref{42})
in the spinor representation
we need the two matrices
\begin{eqnarray*}
  U\gamma^0 \hat{P}_2 U^{-1}&=&
  E_{2}\,
  \left(
  \begin{array}{cc}
      1+\sigma_z & 0 \\
      0& 1-\sigma_z
   \end{array}
  \right),\\
  U\gamma^0\hat{e}_\perp U^{-1} &=&
  \left(
  \begin{array}{c c}
     -{\bf e}_\perp \mbox{\boldmath$\sigma$}_\perp & 0  \\
     0 & {\bf e}_\perp \mbox{\boldmath$\sigma$}_\perp
  \end{array}
  \right) \,.
  \label{aa8}
\end{eqnarray*}

\end{document}